\documentclass{ws-jai}
\usepackage[flushleft]{threeparttable}

\usepackage{aas_macros}
\usepackage{caption}
\usepackage{subcaption}
\usepackage{hyperref}
\usepackage{enumitem}
\usepackage{siunitx}

\usepackage{xcolor}
\hypersetup{
    colorlinks,
    linkcolor={red!50!black},
    citecolor={blue!50!black},
    urlcolor={blue!80!black}
}

\begin{document}

\catchline{}{}{}{}{} 

\markboth{Naveen Dukiya}{Astrometric and photometric standard candidates for the upcoming 4-m ILMT survey}

\title{Astrometric and photometric standard candidates for the upcoming 4-m International Liquid Mirror Telescope survey}

\author{Naveen Dukiya$^{1}$, Kuntal Misra$^{1}$, Bikram Pradhan$^{2}$, Vibhore Negi$^{1, 3}$, Bhavya Ailawadhi$^{1, 3}$, Brajesh Kumar$^{1}$, Paul Hickson$^{4}$, Jean Surdej$^{5}$}

\address{
$^{1}$Aryabhatta Research Institute of observational sciencES, Manora Peak, Nainital, 263001, India\\
$^{2}$Indian Space Research Organization (ISRO) Headquarters, Bengaluru, Karnataka, 560094, India\\
$^{3}$Department of Physics, Deen Dayal Upadhyay Gorakhpur University, Gorakhpur 273009, India\\
$^{4}$Department of Physics and Astronomy, University of British Columbia, 6224 Agricultural Road, Vancouver, BC V6T 1Z1, Canada \\
$^{5}$Space sciences, Technologies and Astrophysics Research (STAR) Institute, Universit\'{e} de Li\`{e}ge, All\'{e}e du 6 Ao\^{u}t 19c, 4000 Li\`{e}ge, Belgium
}

\maketitle

\corres{$^{1}$ndukiya@aries.res.in, ndookia@gmail.com}

\begin{history}
\received{(to be inserted by publisher)};
\revised{(to be inserted by publisher)};
\accepted{(to be inserted by publisher)};
\end{history}

\begin{abstract}
The International Liquid Mirror Telescope (ILMT) is a 4-meter class survey telescope that has recently achieved first light and is expected to swing into full operations by 1st January 2023. It scans the sky in a fixed \ang{;22;} wide strip centered at the declination of \ang{+29;21;41} and works in \emph{Time Delay Integration (TDI)} mode. We present a full catalog of sources in the ILMT strip that can serve as astrometric calibrators. The characteristics of the sources for astrometric calibration are extracted from \textit{Gaia} EDR3 as it provides a very precise measurement of astrometric properties such as RA ($\alpha$), Dec ($\delta$), parallax ({$\pi$}), and proper motions ($\mu_{\alpha^{*}}$ \& $\mu_{\delta}$). We have crossmatched the \textit{Gaia} EDR3 with SDSS DR17 and PanSTARRS-1 (PS1) and supplemented the catalog with apparent magnitudes of these sources in $g, r$, and $i$ filters. We also present a catalog of spectroscopically confirmed white dwarfs with SDSS magnitudes that may serve as photometric calibrators. The catalogs generated are stored in a SQLite database for query-based access. We also report the offsets in equatorial positions compared to \textit{Gaia} for an astrometrically calibrated TDI frame observed with the ILMT.
\end{abstract}


\section{Introduction}
\label{introduction}

Optical survey telescopes have been an integral part of modern-day astronomy for quite some time. Unlike targeted observations, survey telescopes catalog a large number of celestial objects. Such catalogs can often negate the need for dedicated observations of a source of interest, or at the very least can provide useful information before follow-up observations are required for a more detailed study. The Sloan Digital Sky Survey (SDSS) maps the sky with deep photometry and spectroscopy; the first part of the Panoramic Survey Telescope and Rapid Response System (PanSTARRS1 or PS1) is the world's leading Near-Earth object discovery telescope and also provides deep photometry for $3\pi$ steradian sky; the AAVSO Photometric All-sky survey (APASS) aims to map all of the sky in 8 different filters; The Zwicky Transient Facility (ZTF) focuses on time-domain science. In addition to ground-based surveys, there have been many space-based optical surveys as well. The Transiting Exoplanet Survey Satellite (TESS) searches the sky for exoplanets; the Near Earth Object Surveillance Satellite (NEOSSat) detects Near Earth Objects; the \textit{Gaia} satellite aims at mapping the spatial and velocity distributions of the Milky way with very high astrometric accuracy.

Liquid Mirror Telescopes (LMTs) consist of a class of telescopes in which the primary mirror is made of a thin layer of liquid mercury. The mercury is put in a disk and a parabolic liquid mirror is achieved by spinning the disk at a uniform angular velocity. Due to their working principle, LMTs can only be pointed towards the zenith. This property makes them inadequate for targeted observation of a specific source unless it happens to be in the field of view of the telescope pointing at the zenith. Despite their shortcomings, LMTs have some notable advantages that make them useful for conducting sky surveys, over traditional telescopes. Indeed, the quality of the recorded observations is optimal at the zenith since both the seeing and the transparency are the best there, at all times. Moreover, the observing efficiency is very high. As the telescope is pointing toward the zenith and the same strip of the sky is observed each night, no time is lost in slewing the telescope as well as in acquisition of the field, readout time, etc. While observing at zenith, the flat fielding and CCD image de-fringing are much more accurate than during classical observations since the images are actually formed by averaging the signal over entire CCD columns (in the direction of the scan). As compared to traditional telescopes, LMTs are an order of magnitude cheaper to install.
LMTs scan almost the same strip of the sky over and over again every night, this makes them particularly useful in studies of transients. Additionally, co-adding the images from the same patch of sky can be helpful for deep-sky exploration.

The 4-m International Liquid Mirror Telescope (ILMT) is the newest of this  class of telescopes situated at Devasthal, Uttarakhand, India at an altitude of $\sim2450m$ above mean sea level to take advantage of the sub-arcsecond seeing available at the site \citep{sagar2000, stalin2001,Sagar2012}. ILMT saw its first light in the commissioning phase that took place during April--May of 2022 and will resume regular observations from October 2022 after the monsoon break. ILMT has a Field-of-View (FoV) of \ang{;22;} and it will repeatedly observe $\sim 40$ square degrees of sky per night in a strip of the same width (\ang{;22;}) centered on the declination that is equal to the latitude of the site, i.e. \ang{29;21;41}N. Throughout the year, ILMT will survey around $\sim 120$ square degrees of sky and it will perform observations in $g', r'$, and $i'$ filters based on the Sloan Digital Sky Survey photometric system for observations \citep{surdej2018}. 

The ILMT will perform observations by operating the CCD detector in Time Delay Integration (TDI) mode. Due to the Earth's rotation, the sky will move across the detector at the sidereal rate. TDI observation is achieved by matching the parallel charge transfer rate of the CCD with the sidereal rate. This technique provides an effective integration time of $\sim 102$ s (the time it takes a source to move across the detector)  for ILMT. The images observed in this way will be in pixel coordinates. In order to convert the pixel coordinates to equatorial ones, it is essential to have a catalog of sources with well-defined positions which will help in aligning the observed frames with a standard frame such as the J2000 or the ICRS. These sources should also have very low parallax and proper motions so deviations in their positions during the run of a survey is very small. The primary aim of our work is to prepare a catalog of sources that can be used as astrometric calibrators for the data that will be generated with the ILMT in the forthcoming operations. We have used the data from the European Space Agency's {\it Gaia} satellite, which features unparalleled astrometric accuracy, and crossmatched it with data from SDSS DR17 and PS1 DR1 to create a database of astrometric and photometric calibrators. Our aim for crossmatching \textit{Gaia} EDR3 with SDSS and PS1 is to exclude spurious sources from the catalog and have $g, r$, and $i$ magnitudes of all these objects.

The paper is structured as follows: in section \ref{sec:sky_coverage}, we describe the variation in the sky-coverage of the ILMT over the course of its expected runtime.
Section \ref{sec:astrometric_catalog} is divided into different sub-sections providing details of the astrometric calibration catalog such as the description of the data used (section \ref{sec:data_used}), the methodology followed to arrive at the final catalog (section \ref{sec:methodology}). The crossmatching results with SDSS and PS1 are discussed in section \ref{sec:xm_results} and the characteristics of the final catalog are presented in section \ref{sec:final_catalog}. In section \ref{sec:quasar_catalog}, we have drawn a comparison between this work and \citet{Mandal_2020}. The applications of this catalog are described in section \ref{sec:applications}. The results of this work are summarised in section \ref{sec:summary}.

\section{ILMT Sky Coverage} \label{sec:sky_coverage}

Ground-based telescopes observe in a reference frame that is defined by projecting the Earth's equator and poles onto the celestial sphere, and the vernal equinox as a starting point for the celestial longitude (right ascension). This frame, however, is not fixed and varies with respect to the very distant celestial objects due to the movement of the Earth's rotation axis, caused by the gravitational influence of bodies in the solar system, mainly the Sun and the Moon. These effects can be broken down as (i) Precession - which shows variations with a timescale of $\sim$26,000 years and (ii) Nutation - the more short-term variation over a period of $\sim$18 years and amplitude of about 9 arcseconds. Standard frames of reference such as FK5 and ICRS consider a frame fixed in time defined with respect to very distant quasars with negligible proper motion.

\begin{figure}[htbp]
    \centering
    \includegraphics[width=5in]{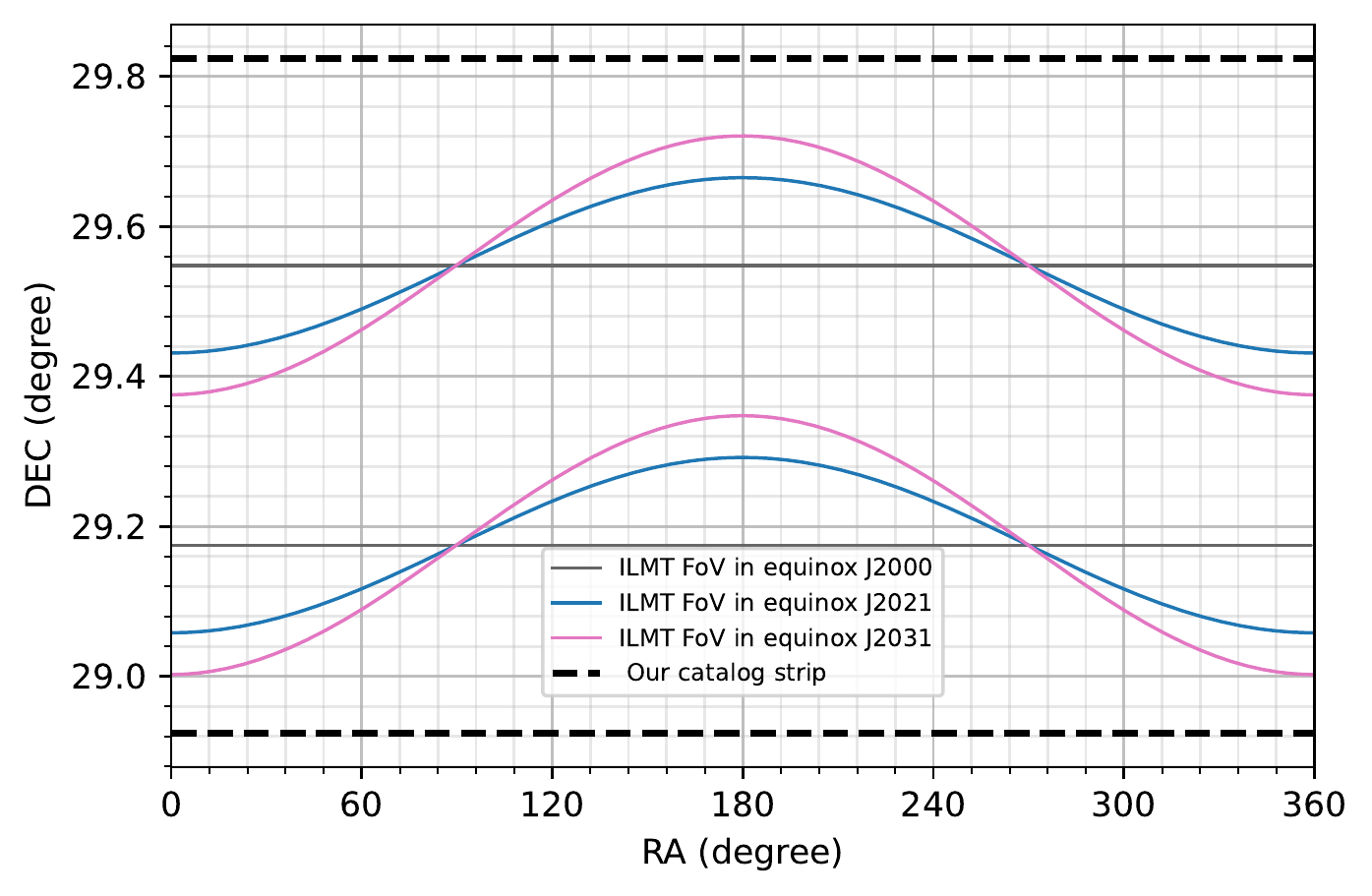}
    \caption{FoV of ILMT for different equinoxes in ICRS frame coordinates (gray, blue, pink). The black dashed line shows the extent of {\it Gaia} sources we have chosen for our catalog.}
    \label{fig:ilmt_precession}
\end{figure}

The variation in the Earth's reference frame means that the area of the sky that ILMT observes will change over time as shown in Figure \ref{fig:ilmt_precession}. To make a catalog of sources that will populate the FoV of ILMT throughout its expected runtime of 10 years, we choose a strip that is \ang{;54;} wide in the ICRS frame coordinates. For this calculation, we have only considered the effects of precession as the effects of nutation, and aberration (the periodic shift in the apparent position of objects in the sky due to the orbital velocity of the Earth), are negligible compared to the FoV of the telescope.

\section{Astrometric Calibration catalog}
\label{sec:astrometric_catalog}

\subsection{Data used}\label{sec:data_used}
To create a catalog suitable for astrometric calibration of a large survey, we have chosen to make use of the early third data release (EDR3) of the \textit{Gaia} satellite \citep{gaia2021} by the European Space Agency (ESA). \textit{Gaia} is a space-based mission and its primary aim is to map the three-dimensional spatial and three-dimensional velocity distributions of astronomical sources \cite{gaia_mission} and it features very precise astrometric solutions of the same. It has at least a 5-parameter astrometric solution - right ascension (RA), declination (Dec), parallax, and the two components of proper motion for about 1.468 billion sources \citep{lindegren2021}. It contains \textit{Gaia} G magnitudes of about 1.806 billion sources and $G_{BP}$ and $G_{RP}$ magnitudes of about 1.542 billion and 1.555 billion sources, respectively \citep{riello2021}.

We have only chosen sources that have low proper motions ($\mu < 20$ mas/yr) and parallax ($\delta < 10$ mas) values. These thresholds are taken from the quasar catalog of \citet{Mandal_2020} and represent the typical upper limit of these parameters for quasars. Additionally, we need to consider the \textit{astrometric excess noise} ($\epsilon_i$) which quantifies the disagreement between the observation and the best-fitting astrometric model adopted by \textit{Gaia}. The significance of this noise is given by the parameter \textit{astrometric excess noise significance} ($D$) in the \textit{Gaia} catalog. As suggested by \citet{lindegren2021}, we selected sources that have $D<2$. These filters were performed in the \textit{Gaia} Data Archive itself using the following ADQL query -

\begin{center}
\begin{verbatim}
    WHERE dec between 28.9239 and 29.8238 
    AND ABS(parallax) < 10
    AND pm < 20
    AND astrometric_excess_noise_sig < 2
\end{verbatim}
\end{center}

In the $54'$ strip, there are $\sim 7.62$ million \textit{Gaia} objects. After applying the mentioned conditions on $\mu$, $\delta$, and $D$, we were left with $\sim 5.46$ million \textit{Gaia} objects. 
These sources will define a self-consistent frame of reference that can be used for the astrometric calibration of fields observed by ILMT.

We have crossmatched the \textit{Gaia} sources with the 17th data release (DR17;  \citealt{sdssdr17}) of the SDSS. It is the final data release for the fourth phase of the SDSS survey (SDSS-IV) \citep{sdssIV}. SDSS is a very ambitious and detailed survey that has been primarily mapping out the northern galactic hemisphere since the year 2000 using a 2.5m modified Ritchey-Chrétien telescope \citep{sdss_telescope} situated at Apache Point Observatory, New Mexico, USA. SDSS provides photometry in $u$, $g$, $r$, $i$, $z$ bands \cite{sdss_photometric_system} upto a median $5\sigma$ depth of 22.15, 23.13, 22.70, 22.20, 20.71 mag, respectively. There are $\sim 6.37$ million sources in the $54'$ strip in SDSS DR17. It also contains spectroscopic data for about $\sim 1\%$ of these objects.

SDSS mostly covers the sky at high galactic latitudes therefore, there are some gaps in the survey inside the region of our interest, as seen in section \ref{fig:object_density_sdss}. To overcome this issue, we supplement our catalog with data products generated by crossmatching \textit{Gaia} EDR3 with the first part of the Panoramic Survey Telescope and Rapid Response System (Pan-STARRS1 or PS1) Data Release 1 (DR1; \citealt{panstarrs_survey}). PS1 made use of a 1.8m telescope situated at Haleakala Observatories in Hawaii to survey the sky in $g$, $r$, $i$, $z$, and $y$ bands north of declination \ang{-30;;}. The $5\sigma$ photometric depth in $g$, $r$, $i$, $z$, $y$ bands is 23.3, 23.2, 23.1, 22.3, 21.3 mag respectively for stacked images. \texttt{Casjobs}\footnote{\url{http://mastweb.stsci.edu/ps1casjobs/}} was used to extract the objects that have more than 1 detection in the survey from \emph{ObjectThin} and \emph{MeanObject} tables (see \citealt{panstarrs_data_products}).

SDSS and PS1 both are ground-based surveys and deeper than \textit{Gaia} in photometric depth and hence can be crossmatched using the same methodology. In addition to that, both of these list the photometric magnitudes in $g$, $r$, and $i$ bands, which can be converted to $g'$, $r'$, and $i'$ filter magnitudes used by ILMT\footnote{\url{http://classic.sdss.org/dr5/algorithms/fluxcal.html}}. From both these catalogs, we only choose sources that have positional errors less than \ang{;;2}.

In addition to that, we have also used the data from the Miliquas catalog and the Montreal White Dwarf database to derive additional data products. Details of which are listed in sections \ref{sec:quasar_catalog} and \ref{sec:photometic_calibration_catalog} respectively.

\subsection{Methodology}
\label{sec:methodology}

To create a catalog that is suitable for the astrometric calibration of the fields observed by ILMT, we selected all the \textit{Gaia} sources having low proper motion and parallax values and insignificant astrometric noise in a \ang{;54;} wide strip centered around the declination of \ang{29;21;41} N. These sources were then crossmatched with SDSS DR17 and PS1 DR1. Our motivation for crossmatching the \textit{Gaia} sources with the other catalogs is the following: 

\begin{enumerate}[label=(\roman*), leftmargin=2em]
    \item Since the observation epoch of distinct surveys is different, a crossmatch between these would eliminate the spurious sources that might have been observed by one of the surveys.
    \item Both SDSS and PS1 have observations in $g$, $r$, and $i$ bands, which can be converted to $g'$, $r'$ and $i'$ system used by ILMT. Photometric magnitudes in these bands can also serve as an additional constraint when calibrating the ILMT fields for astrometry.
    \item Additional data products like a photometric calibration catalog can be derived based on the $g$, $r$, and $i$ band photometry.
\end{enumerate}

\noindent
Since both SDSS and PS1 are deeper surveys than \textit{Gaia}, we have chosen \textit{Gaia} as the leading catalog and SDSS or PS1 as the secondary one. This approach is consistent with crossmatching of \textit{Gaia} DR2 with multiple other catalogs by \citet{Marrese_2019}. In the crossmatching procedure, we have only made use of the astrometric parameters listed in the catalogs and considered angular proximity of \ang{;;2} as the limiting criteria for finding a match. We choose this limiting criterion because most of the objects in the ground-based surveys  are observed with seeing less than \ang{;;2}. We then further refine our crossmatch with the Good-neighbour criteria explained in \citet{pineau}, the details of which are described below. The procedure to generate the final catalog can be summed up by the following steps: \\

\noindent
1. We select all the sources from \textit{Gaia} EDR3 lying between the declination range of \ang{28.9238} and \ang{29.8239}, i.e. within the \ang{;54;} wide strip centered at the declination of \ang{29.361}. From there, we drop sources that have high proper motion, high parallax, and significant astrometric noise. All these steps are performed using an ADQL query in the \textit{Gaia} data archive as explained in section \ref{sec:data_used}. At this step, we were left with $\sim 5.46$ million \textit{Gaia} sources. 
Similarly, we select all the sources from the secondary catalog within this region and sort them by declination. In the narrow regions of \ang{;;3} around the Right Ascension value of \ang{0} and \ang{360}, we pad the secondary catalog with sources that have RA$<$\ang{0} and RA$>$\ang{360} using sources that have equivalent RA. This is done in order to find the crossmatch for sources that lie near RA of \ang{0} and \ang{360} in the leading catalog.\\

\noindent
2. To find crossmatch for a particular \textit{Gaia} source (or primary object), we select all the sources in the secondary catalog that lie in a square region of half-width \ang{;;2.4} centered on the primary object. Then, we propagate the astrometric parameters and covariance matrix of the \textit{Gaia} sources to the individual observation epoch of all the secondary catalog sources in that square region. To achieve this we use the \texttt{PyGaia}\footnote{\url{https://github.com/agabrown/PyGaia}} library which implements the treatment described by \citet{1997ESA} (1997), Vol. 1, Sect. 1.5. We then calculate the angular distances between the new positions of the primary object and the corresponding secondary objects using \emph{Vicenty Formula} - 
\[ {\displaystyle \Delta \theta =\tan^{-1} \left( {\frac {\sqrt {\left(\cos \delta _{2}\sin(\Delta RA )\right)^{2}+\left(\cos \delta _{1}\sin \delta _{2}-\sin \delta _{1}\cos \delta _{2}\cos(\Delta RA )\right)^{2}}}{\sin \delta _{1}\sin \delta _{2}+\cos \delta _{1}\cos \delta _{2}\cos(\Delta RA )}} \right) }\]
where $\delta_1$ and $\delta_2$ are the declinations and $\Delta RA$ is the difference in the Right Ascension of both sources.\\
Finally, we select the source that is closest to the primary object up to an angular separation of \ang{;;2} as a potential counterpart.\\

\noindent
3. In order to further refine the crossmatch, we apply the good-neighbor criteria described in \citet{pineau}. In this approach, we use the positional uncertainties in equatorial coordinates of these sources to decide whether the potential counterpart provides a good match to the primary object or not. Positional errors in astronomy are described by error ellipses around the true value. We take the error ellipses in the equatorial coordinates and project them into the tangent plane centered on the primary object, with the $x$-axis toward the potential counterpart. Errors are then interpreted as 2D gaussians:
\begin{align}
    &N_p(x, y; \sigma_{x_p}^2, \sigma_{y_p}^2, \rho_p\sigma_{x_p}\sigma_{y_p}) \nonumber \\
    &N_s(x-d, y; \sigma_{x_s}^2, \sigma_{y_s}^2, \rho_s\sigma_{x_s}\sigma_{y_s})    
\end{align}
Here, $\sigma_x$ and $\sigma_y$ are positional uncertainties in the $x$ and $y$ directions respectively and $\rho$ is the correlation. The subscripts $p$ and $s$ denotes the primary and secondary object. $d$ is the distance between these objects.

The density of probability that the two objects are the same source is given by the convolution of these two distributions: 

\begin{equation}\label{eqn:conv_distribution}
    N_c(x-d, y; \sigma_{x_c}^2, \sigma_{y_c}^2, \rho_c\sigma_{x_c}\sigma_{y_c})        
\end{equation}
where, 
$\sigma_{x_c}^2 = \sigma_{x_p}^2 + \sigma_{x_s}^2, \,
    \sigma_{y_c}^2 = \sigma_{y_p}^2 + \sigma_{y_s}^2, \, \mathrm{and} \, \,\,
    \rho_c\sigma_{x_c}\sigma_{y_c} = \rho_p\sigma_{x_p}\sigma_{y_p} + \rho_s\sigma_{x_s}\sigma_{y_s}
$.\\

If the secondary object is a true counterpart of the primary object, then it will fall inside the confidence region with a probability $\gamma$, defined by the ellipse
\begin{equation}\label{eqn:conv_vcm}
    \begin{pmatrix}
        x \\ y
    \end{pmatrix}^T
    \begin{pmatrix}
        \sigma_{x_c}^2 & \rho_c\sigma_{x_c}\sigma_{y_c} \\
        \rho_c\sigma_{x_c}\sigma_{y_c} & \sigma_{y_c}^2
    \end{pmatrix}
    \begin{pmatrix}
        x \\ y
    \end{pmatrix} = k_{\gamma}^2
\end{equation}
$k_{\gamma}$ is known as Mahalanobis distance and $k_{\gamma}^2$ has a $\chi^2$ distribution with 2 degrees of freedom.\\

In the catalogs used in the present work, the uncertainties in positions are given in equatorial coordinates ($\sigma_{\alpha}$, $\sigma_{\delta}$, and $\rho_{\alpha\delta}$). The different column names for these quantities based on the catalog are given in Table \ref{tab:astrometric_parameters_cat}. We assume the correlation between RA and Dec to be zero when it is not provided. Using these quantities we can define the variance-covariance matrix as 

\begin{equation}\label{eqn:vcm_ra_dec}
    V_0 =
    \begin{pmatrix}
        \sigma_{\alpha}^2 & \rho_{\alpha\delta}\sigma_{\alpha}\sigma_{\delta} \\
        \rho_{\alpha\delta}\sigma_{\alpha}\sigma_{\delta} & \sigma_{\delta}^2.
    \end{pmatrix}
\end{equation}

In order to convert this variance-covariance matrix to the tangent frame, we need to rotate Equation \ref{eqn:vcm_ra_dec} by  an amount defined by the position angle of the potential counterpart source.

\begin{equation}\label{eqn:vcm_tangent_ra_dec}
    V_t =
    \begin{pmatrix}
        \sigma_{x}^2 & \rho\sigma_{x}\sigma_{y} \\
        \rho\sigma_{x}\sigma_{y} & \sigma_{y}^2
    \end{pmatrix}
    =
    R(\Omega)^T
    \begin{pmatrix}
        \sigma_{\alpha}^2 & \rho_{\alpha\delta}\sigma_{\alpha}\sigma_{\delta} \\
        \rho_{\alpha\delta}\sigma_{\alpha}\sigma_{\delta} & \sigma_{\delta}^2
    \end{pmatrix}
    R(\Omega)
\end{equation}

where $R(\Omega)$ is the rotation matrix defined by

\begin{equation*}
    R(\Omega) =
    \begin{pmatrix}
        \cos\Omega \;&\; -\sin\Omega \\
        \sin\Omega \;&\; \cos\Omega
    \end{pmatrix}.
\end{equation*}

The convolved variance-covariance matrix in Equation \ref{eqn:conv_vcm}, now can be represented by the sum of the individual variance-covariance matrices. In the frame we have chosen, $x=d$ and $y=0$, therefore, the selection criteria can be represented as: 

\begin{equation}
    \frac{d}{\sigma_{x_c} \sqrt{(1 - \rho_c^2)}} \leq k_{\gamma}.
\end{equation}

In this work, we have adopted a confidence level of 99.7\%, which corresponds to $k_{\gamma}$ =$\sqrt{11.8290}$. It means if a source-counterpart pair does not satisfy this criterion, we can say with 99.7\% certainty that it is not an actual counterpart.\\

\begin{table}[hb]
    \centering
    \caption{Astrometric parameters for the catalogs used in this work.}
    \begin{tabular}{c|c|c|c} 
     & \textit{Gaia} EDR3 & SDSS DR17 & PS1 DR1 \\ \hline
    $\sigma_{\alpha}$ & \textit{ra\_error} & \textit{raErr} & \textit{raMeanErr} \\
    $\delta_{\alpha}$ & \textit{dec\_error} & \textit{decErr} & \textit{decMeanErr} \\
    $\rho_{\alpha\delta}$ &\textit{ ra\_dec\_corr} & - & - \\
    \end{tabular}
    \label{tab:astrometric_parameters_cat}
\end{table}

4. After the crossmatch, we only kept sources that have $i$ magnitudes between $16.5$ and $22.0$ $mag$, the bright and faint limit of ILMT, respectively. In addition to that, we have removed \textit{Gaia} sources that were having duplicate matches to the same counterparts in the secondary catalogs. These objects are possible binary systems or closeby sources that were resolved in \textit{Gaia} but not in the secondary catalogs.

\subsection{Results}
\label{sec:xm_results}

\begin{figure}[htbp]
    \centering
    \begin{minipage}[t]{0.47\textwidth}
        \centering
        \includegraphics[width=3.3in]{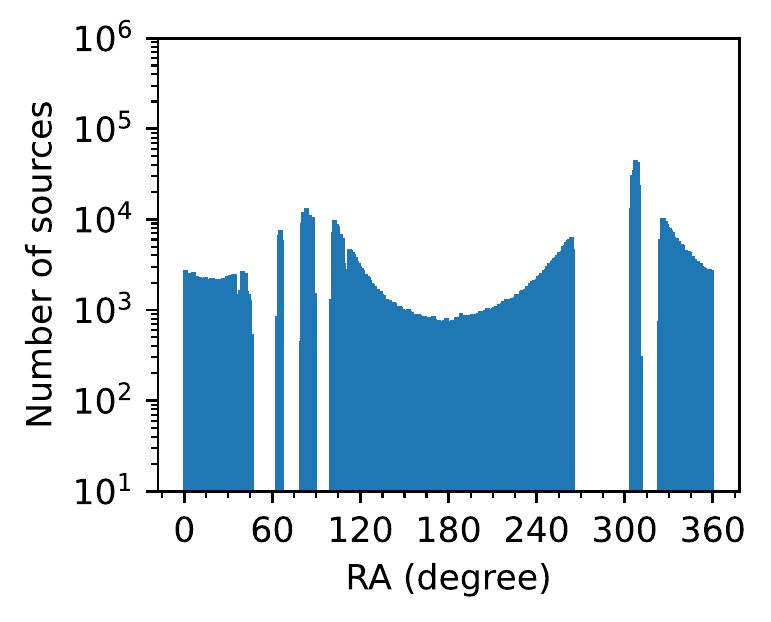}
        \caption{The distribution of \textit{Gaia} sources crossmatched with SDSS at different RA bins of \ang{1}.}
    \label{fig:object_density_sdss}
    \end{minipage}
    \hfill
    \begin{minipage}[t]{0.47\textwidth}
        \centering
        \includegraphics[width=3.2in]{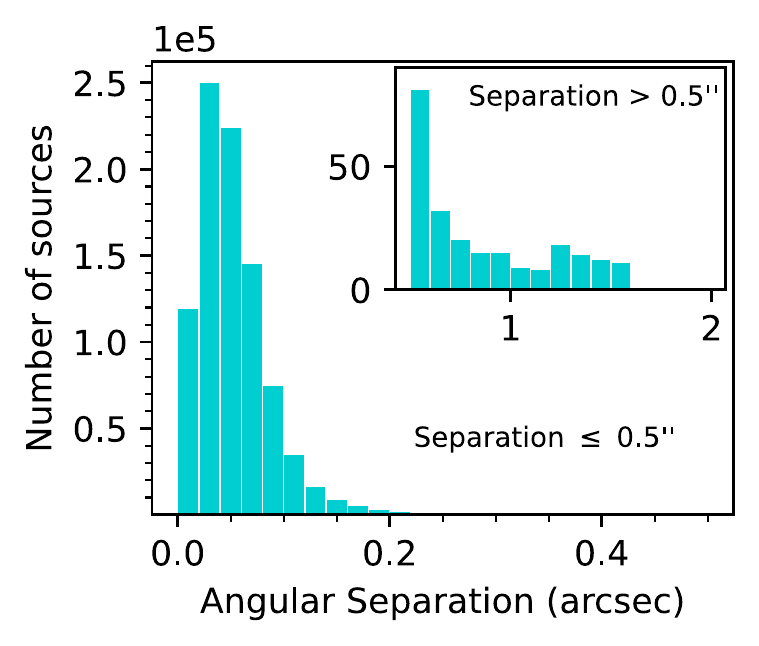}
        \caption{The distribution of angular separation between \textit{Gaia} objects and their SDSS counterpart.}
        \label{fig:ang_sep}
    \end{minipage}
\end{figure}

We were able to find a match for $\sim 1.21$ million \textit{Gaia} sources in SDSS DR17. Since we initially had $\sim 5.46$ million \textit{Gaia} sources, this number may seem low but it is justified by the fact that SDSS has negligible sky coverage in the much denser galactic equatorial region (see Figure \ref{fig:sky_coverage} later for a description). The distribution of crossmatched sources in different RA ranges is shown in Figure \ref{fig:object_density_sdss} and the histogram of the angular separation between crossmatched \textit{Gaia} sources and their counterparts is shown in Figure \ref{fig:ang_sep}.

\begin{figure}
\begin{minipage}[t]{0.47\textwidth}
    \centering
    \includegraphics[width=3.45in]{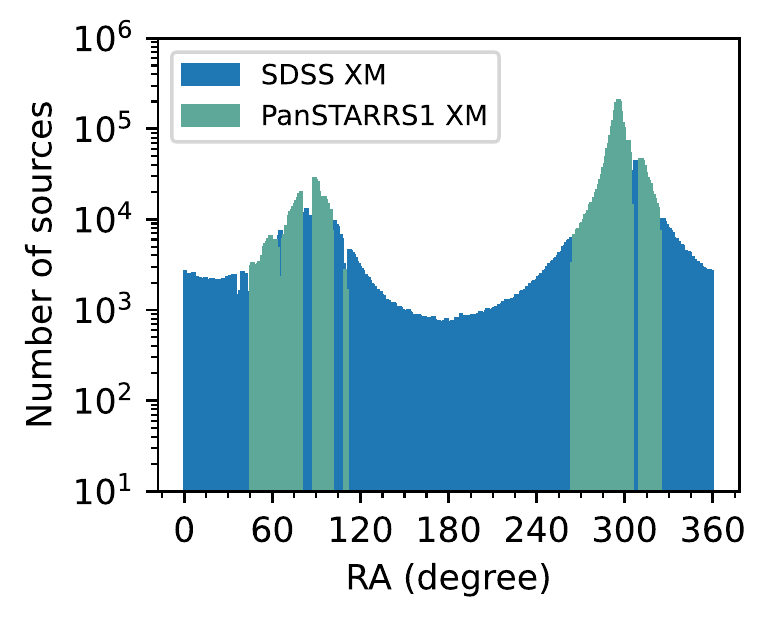}
    \caption{The distribution of \textit{Gaia} sources crossmatched with SDSS and PS1 at different RA bins of \ang{1}.}
    \label{fig:RA_distribution_panstarrs}
\end{minipage}
\hfill
\begin{minipage}[t]{0.47\textwidth}
    \centering
    \includegraphics[width=3.35in]{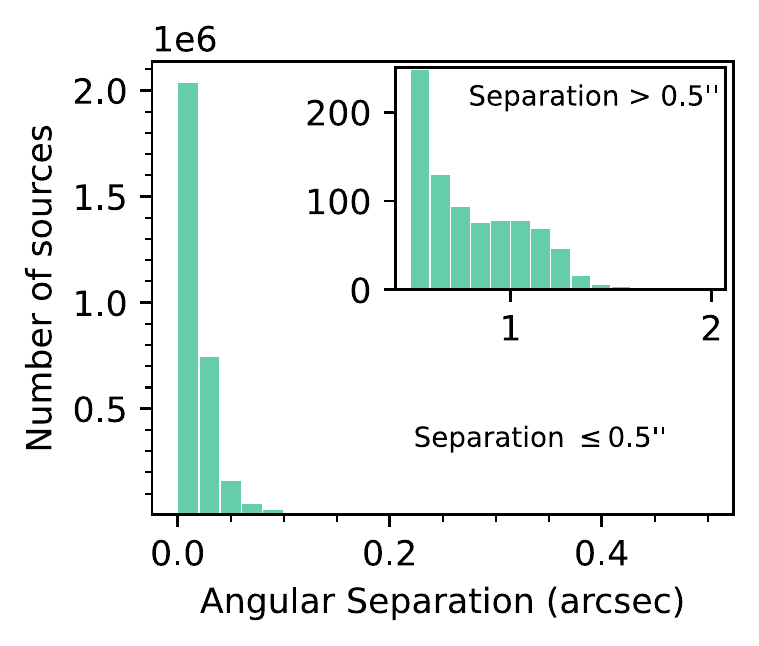}
    \caption{The distribution of angular separation between \textit{Gaia} objects and their PS1 counterpart.}
    \label{fig:ang_sep_panstarrs}
\end{minipage}
\end{figure}

To fill the gaps in regions where little to no crossmatched sources between \textit{Gaia} and SDSS are present, we supplement the catalog by crossmatching \textit{Gaia} EDR3 with PS1 DR1. This results in additional $\sim 3.34$ million crossmatched sources. The distribution of crossmatched sources in different RA ranges, from both SDSS and PS1, is shown in Figure \ref{fig:RA_distribution_panstarrs}.


\begin{table}[tb]
    \centering
    \caption{Summary of objects at different stages of this work.}
    \begin{tabular}{lc} 
    \toprule
    No. of \textit{Gaia} sources in the initial sample & 7.62 million \\
    sources left after $parallax < 10$ mas filter & 6.61 million \\
    sources left after $pm < 20$ mas/yr filter & 6.44 million \\
    sources left after $D < 2$ filter & 5.46 million \\
    & \\
    No. of sources in SDSS sample & 6.37 million\\
    with $positional \, errors < 2''$ & 6.34 million\\
    No. of crossmatches & 0.89 million\\ 
    with spectroscopic observation & 11.3 thousand
    \\
    & \\
    No. of sources in PS1 sample & 11.55 million \\
    with $positional \, errors < 2''$ & 11.53 million\\ 
    No. of crossmatches & 3.04 million\\ 
    & \\
    Total crossmatches & 3.93 million\\ 
    duplicated sources & 544\\
    with $i$ magnitude between 16.5 and 22 & 3.56 million\\  
    \hline
    \end{tabular}
    \label{tab:gaia_filter}
\end{table}

The histogram of the angular separation between the \textit{Gaia} sources and their PS1 counterparts is given in Figure \ref{fig:ang_sep_panstarrs}. One thing that is immediately noticeable from the figure is that the sources in \textit{Gaia} and PS1 are more closely matched than the sources in \textit{Gaia} and SDSS. This is most likely due to the reason that PS1 DR1 sources that were detected in single epoch images use \textit{Gaia} DR1 as a reference for astrometry. So any systematic deviation between these two surveys is lower. Also, the median error in the position of sources within this strip is $\sim 0.08''$ for SDSS, while it is $\sim 0.01''$ for PS1. For comparison, the median error in positions for {\it Gaia} is $\sim 0.0002''.$
    
\subsection{The Final catalog}
\label{sec:final_catalog}

\begin{figure}[htbp]
    \centering
    \includegraphics{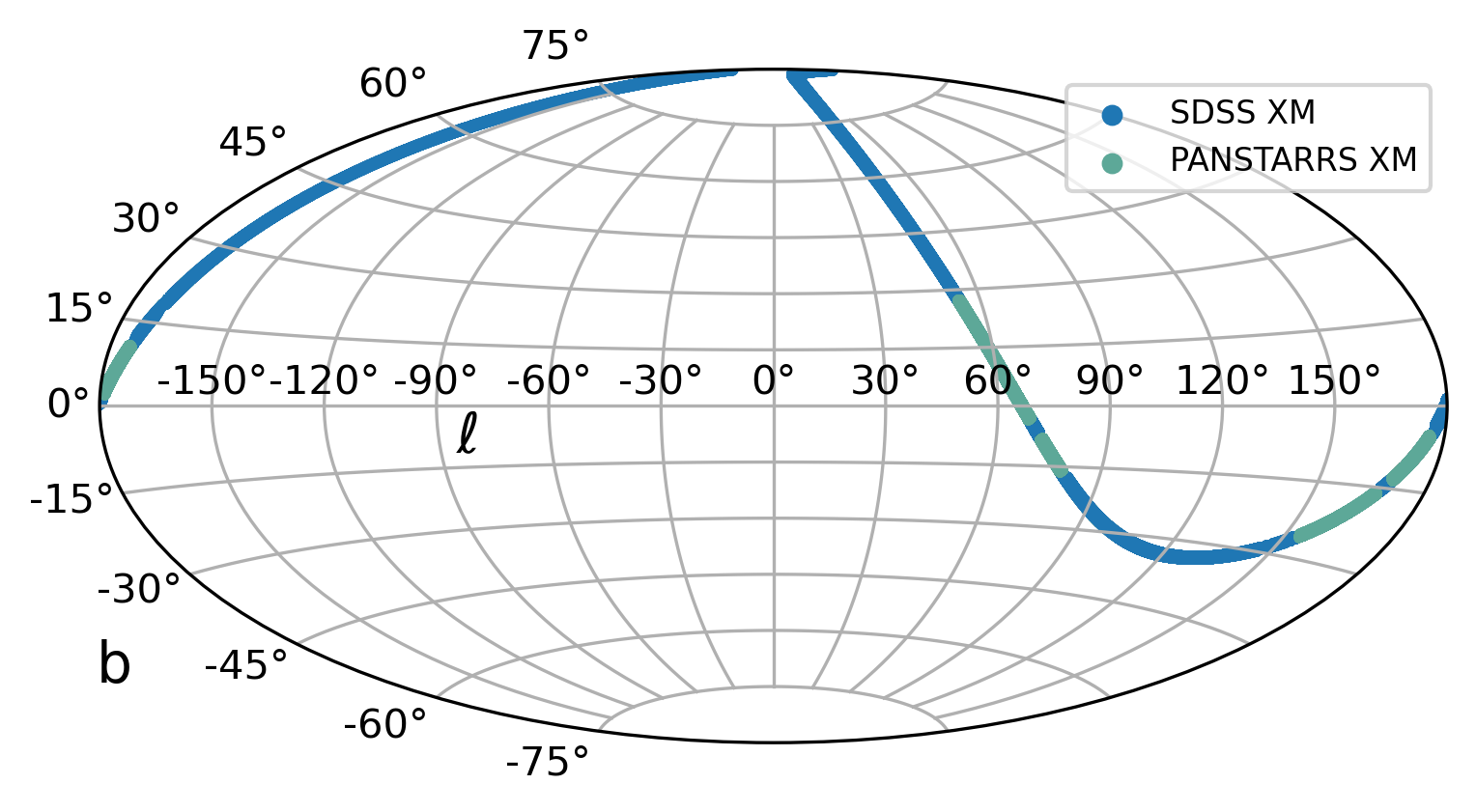}
    \caption{The distribution of sources in the final catalog shown in galactic coordinate system.}
    \label{fig:sky_coverage}
\end{figure}

\begin{figure}[htb]
    \centering
    \includegraphics[width=6in]{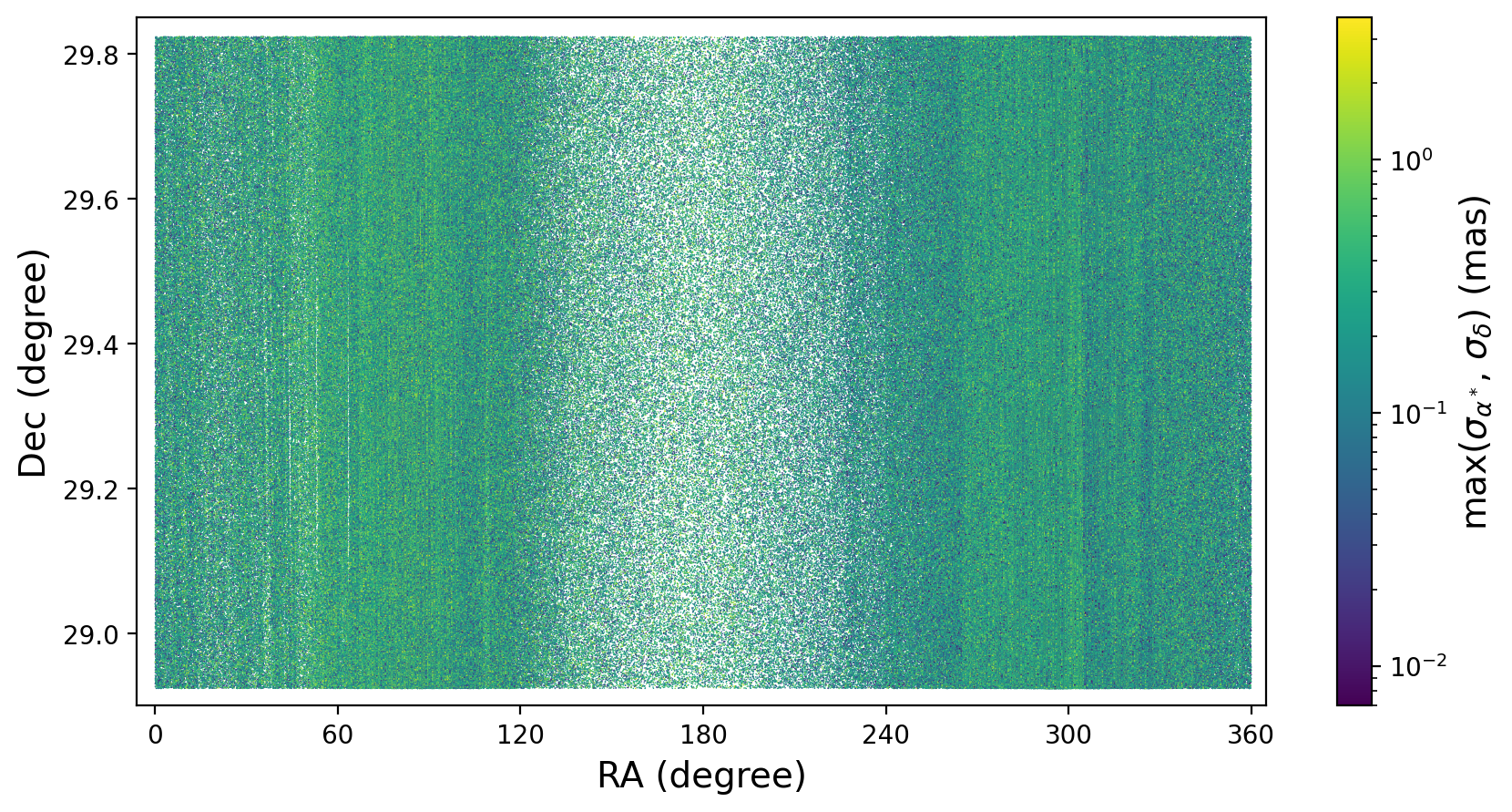} 
    \caption{All crossmatched \textit{Gaia} sources and their associated positional uncertainty max($\sigma_{\alpha^*}, \sigma_{\delta}$) shown in ICRS frame coordinates. The unit of positional uncertainties is miliarcsecond(mas).}
    \label{fig:error_ra_dec}
\end{figure}

\begin{figure}[htb]
    \centering
    \begin{subfigure}{0.5\textwidth}
        \centering
        \includegraphics[width=3.2in]{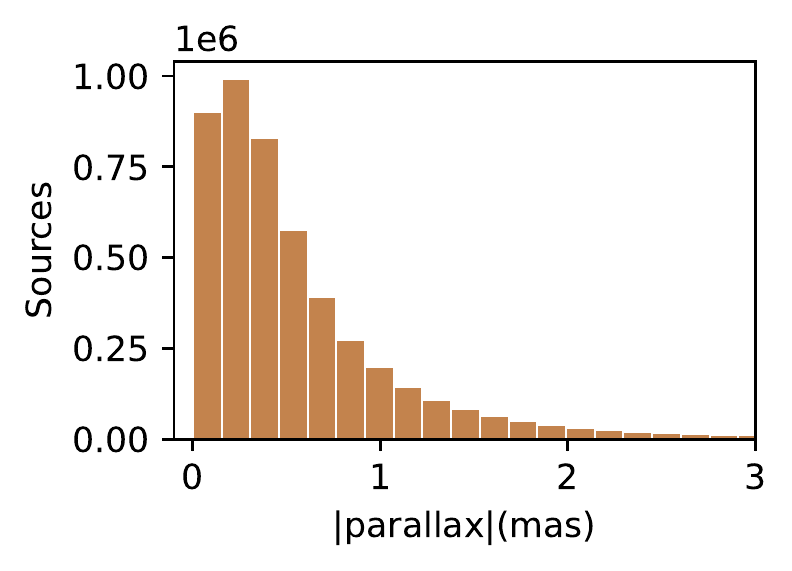} 
    \end{subfigure}%
    \begin{subfigure}{0.5\textwidth}
        \centering
        \includegraphics[width=3in]{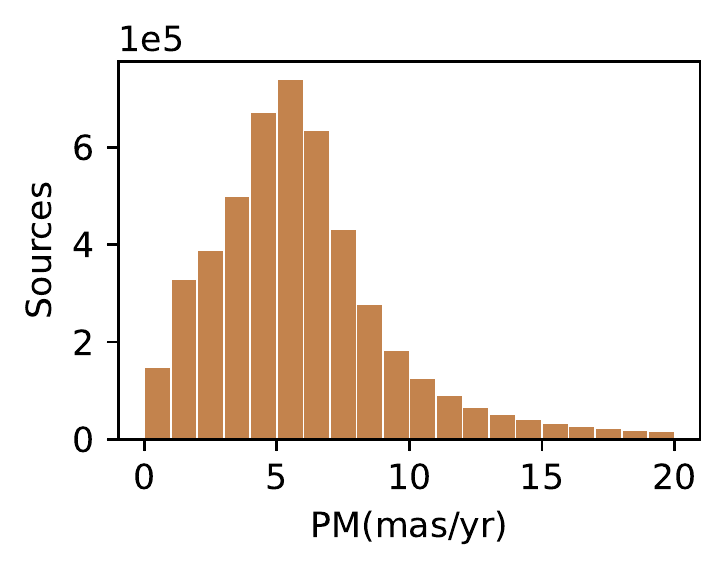}
    \end{subfigure}
    \caption{The distribution of parallax and Proper Motions in the final catalog.}
    \label{fig:parallax_and_pm}
\end{figure}

\begin{figure}[htbp]
    \centering
    \includegraphics[width=5in]{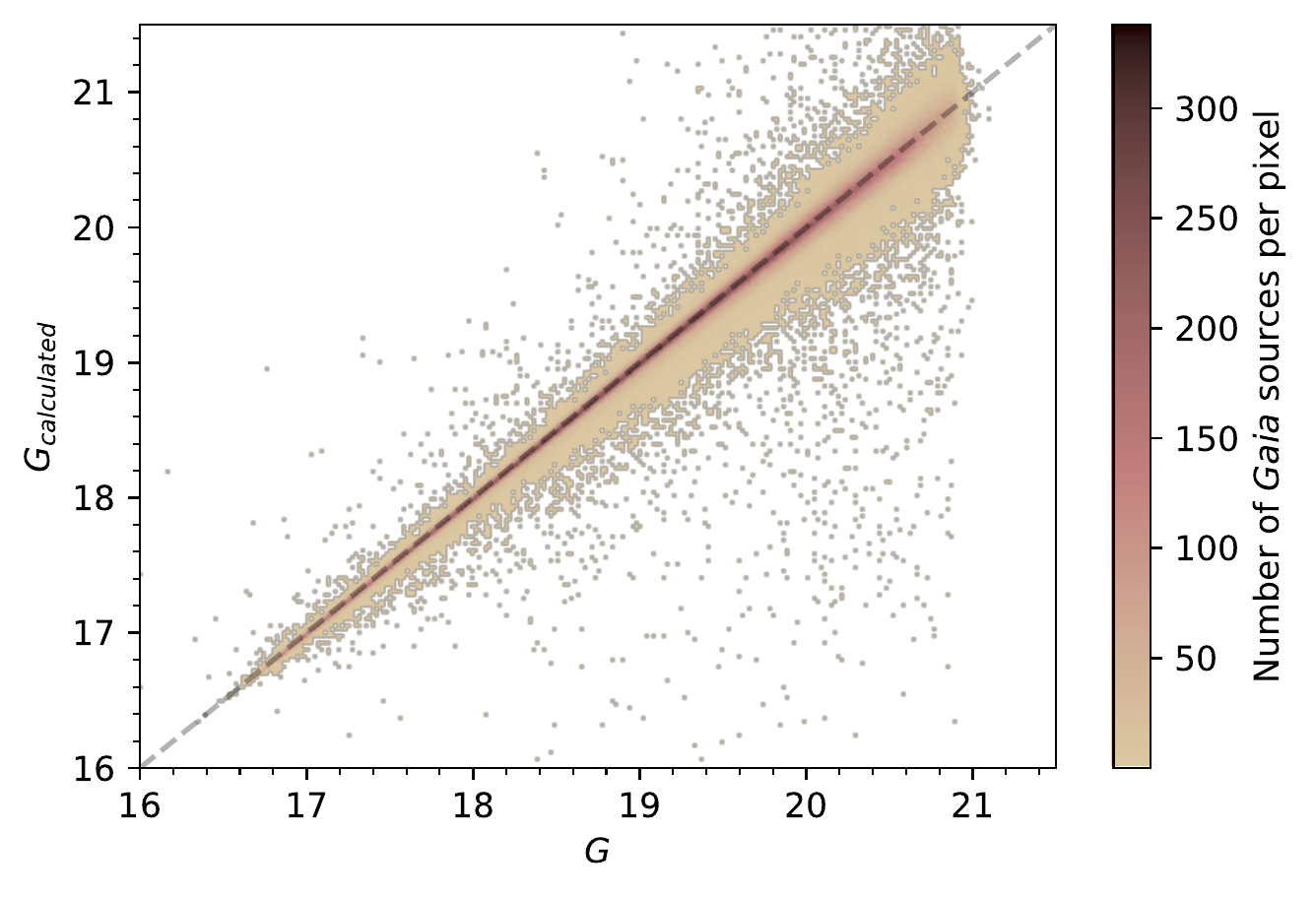}
    \caption{The calculated $G$ magnitude from \citet{jordi} color equations is plotted against \textit{Gaia G} magnitude.}
    \label{fig:mag_comparison}
\end{figure}

The \textit{Gaia} data crossmatched with both SDSS and PS1 will serve as the astrometric calibration catalog. This catalog covers the entire region of the sky that will be observed by the ILMT, providing $250$ sources at the least inside the instantaneous ILMT FoV. The positions in galactic coordinates of the sources present in this catalog are shown in Figure \ref{fig:sky_coverage}. The maximum of uncertainty either in RA or Dec, max($\sigma_{\alpha^*}, \sigma_{\delta}$)\footnote{$\sigma_{\alpha^*} = \sigma_{\alpha}\cos\delta$. Here, $\sigma_{\alpha}$ and $\sigma_{\delta}$ are the errors in RA and Dec respectively.}, for each crossmatched \textit{Gaia} source are shown in figure \ref{fig:error_ra_dec}. The distributions of proper motion and absolute parallaxes of these objects is given in Figure \ref{fig:parallax_and_pm}. The $g$, $r$, $i$ magnitudes of the counterparts were converted to \textit{Gaia} \textit{G} magnitudes using the color equations of Jordi et al.  \cite{jordi}. In Figure \ref{fig:mag_comparison}, the calculated $G$ band magnitude is plotted against the \textit{Gaia} $G$ band magnitude. It shows an expected straight-line trend with some scatter towards the fainter end.

The crossmatched data are stored in a SQLite database that contains the following tables - \textit{crossmatches}, \textit{gaia}, \textit{sdss}, \textit{panstarrs}, and \textit{sdss\_spec}. The \textit{crossmatches} table is meant to be joined with the other  tables so it only contains minimal information about the sources such as \textit{gaia\_id, sdss\_id, angular\_separation, ra, dec} and the other tables contain all the relevant astrometric and photometric information.

\begin{figure}[htbp]
\begin{subfigure}[t]{0.5\textwidth}
    \centering
    \includegraphics[width=3.2in]{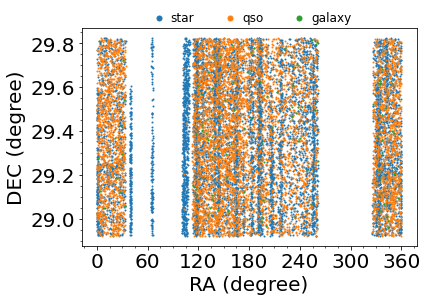}
    \label{fig:spec_obj_sky_coverage}
\end{subfigure}%
\begin{subfigure}[t]{0.5\textwidth}
    \centering
    \includegraphics[width=2.6in]{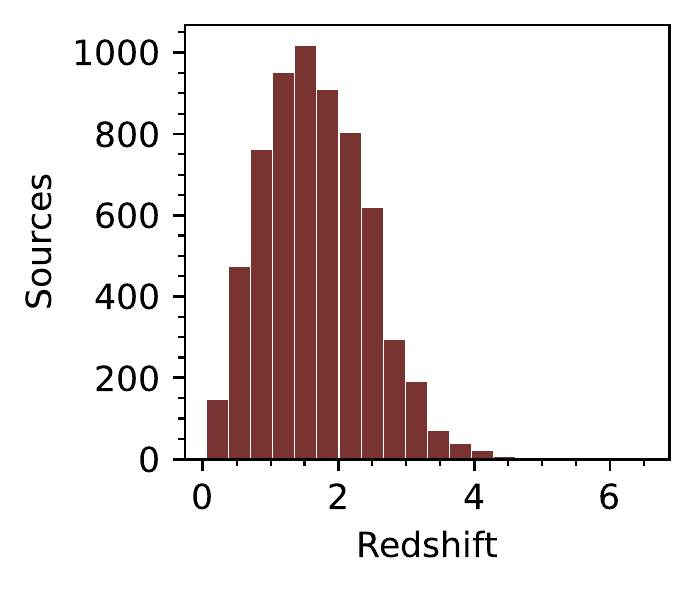}
    \label{fig:spec_obj_redshift}
\end{subfigure}
\caption{\textit{Left figure} shows the sky positions and \textit{Right Figure} shows the redshift distribution of sources from the final catalog that have spectroscopic information available in SDSS.}
\label{fig:spec_obj}
\end{figure}

The SDSS contains spectroscopic information for $\sim 1\%$ of the sources from these crossmatched sources. Spectroscopic information can provide information about the type of these sources. SDSS primarily classifies these sources in 3 categories, \emph{{STAR, QSO}}, and \emph {GALAXY} and then further classifies them into subclasses. The sky-distribution and redshifts of these sources are shown in \ref{fig:spec_obj}. Spectral classification can be useful for photometric calibrations and the study of variability in known QSOs and stars. For this reason, we have opted to include the additional \textit{sdss\_spec} table in the database.

\subsection{Comparison with previous Quasar catalogs}\label{sec:quasar_catalog}
Prior to this work, \citet{Mandal_2020} had presented a catalog of Quasars from the Million-Quasars Database (or Milliquas) in the ILMT strip by crossmatching it with \textit{Gaia} DR2. Quasars show negligible parallax and proper motions and hence are well suited to be astrometric calibrators for surveys.

In their work, they have chosen a \ang{;34;} wide declination strip, taking into account the precession effects from J2019 to J2029. The precession effects from the J2000 reference frame to J2019, however, were not considered. This does mean that some parts of ILMT FoV will not be covered by the quasar catalog. Although our work fulfills the role of a catalog for the astrometric calibration of fields observed by ILMT, it can be argued that such a catalog of Quasars and Quasar candidates is very useful even outside the calibration needs. As discussed by \citet{Mandal_2020}, Quasar variability studies can make use of such a catalog. For this reason, we have reproduced their work using the latest Milliquas catalog \citep{flesch2021} data in a \ang{;54;} wide declination strip and \textit{Gaia} EDR3 data. The characteristics of the sources such as the distribution of parallax and proper motion stay the same as expected. We have found crossmatches for 8029 quasars present in the new Milliquas catalog.

\section{Applications of the catalog}\label{sec:applications}
\subsection{Astrometric calibration of ILMT images}\label{sec:astrometric_calibration_application}
\begin{figure}[htbp]
    \centering
    \includegraphics[height=8cm,width=18cm]{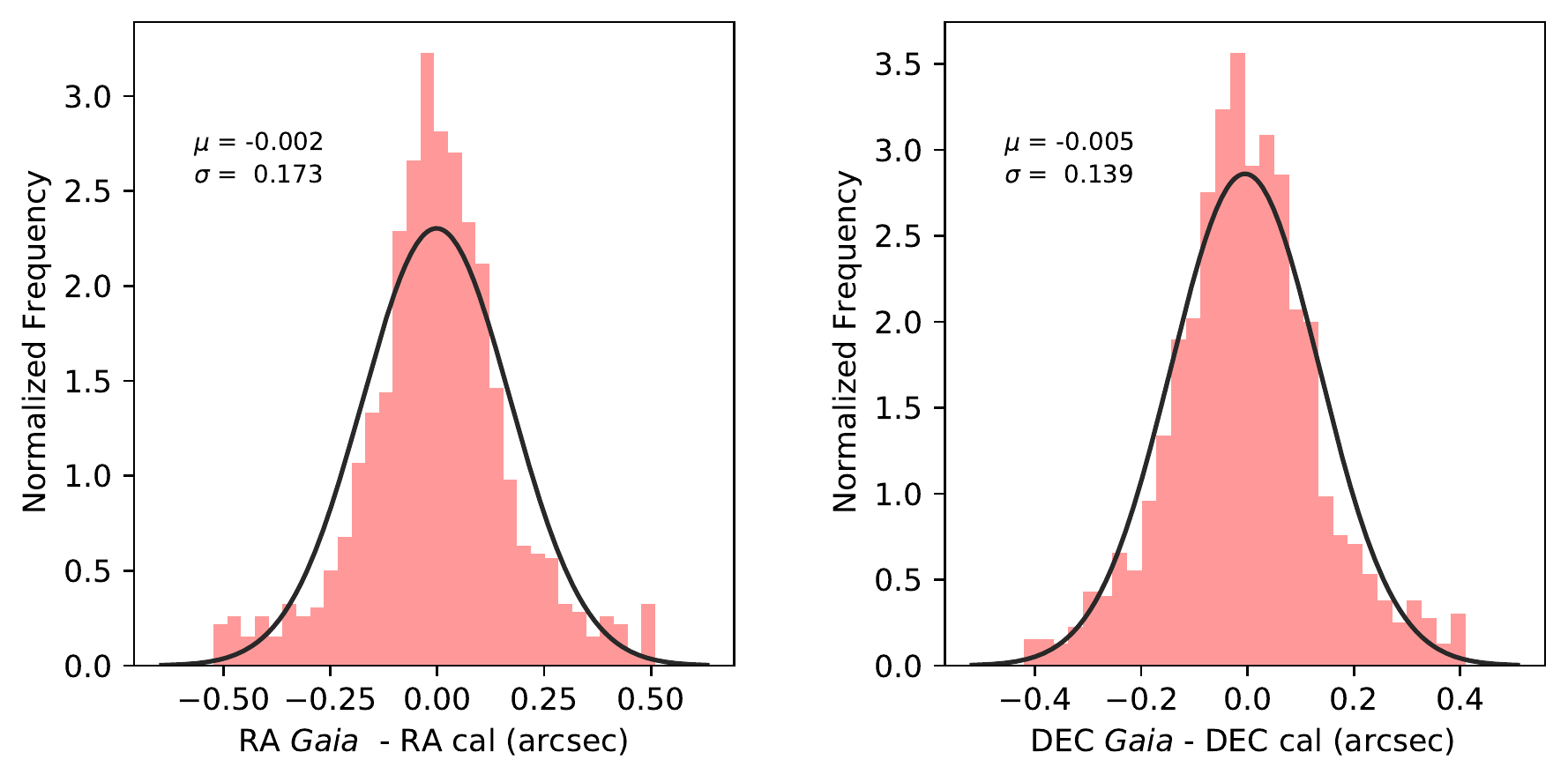}
    \caption{Normalized frequency distribution of the offsets in the calculated astrometric coordinates with respect to the {\it Gaia} coordinates of one TDI frame observed on 23-05-2022 in {\it g} band with the ILMT. The left panel shows the offset distribution in Right Ascensions and the right panel shows the offset distribution in Declinations. The solid black curve indicates the Gaussian function fitted to the distribution with the estimated mean ($\mu$) and standard deviation ($\sigma$) of the best fit given at the top left corner.}
    \label{fig:astrometric_calibration}
\end{figure}

The precise positions of sources from \textit{Gaia} listed in this catalog can be used for the astrometric calibration of the ILMT field. \citet{ILMT_pipeline} describe the data-handling pipeline and the process for the astrometric calibration of ILMT fields in detail. This catalog will be infused into the data handling pipeline of ILMT. After the pre-processing of the images, all the sources will be extracted and registered. A subset of these objects will then be identified using the plate solving engine \texttt{astrometry.net} \citep{astrometry.net}. After this, the entire field will be crossmatched with \textit{Gaia} data to derive the pixel to world coordinate transformation equations. This is where this catalog will be useful. These transformation equations can be derived using this dedicated catalog, which should have little to no spurious sources, to place the ILMT fields in the world coordinates system consistent with \textit{the Gaia} EDR3 frame of reference. With a pixel size of $0.327''$ and median seeing of $\sim 1.1''$ at the Devasthal site, we expect to achieve sub-arcsecond astrometric accuracy in the ILMT survey. To verify this, we have astrometrically calibrated several frames observed with the ILMT, in the commissioning phase, using the method described above. Figure \ref{fig:astrometric_calibration} shows the offset in the calculated equatorial coordinates when comparing the astrometrically calibrated positions of a 17 minute long TDI image (observed on 23-05-2022) to \textit{Gaia}. As can be seen, the standard deviation in the offsets in the Right Ascension and Declination are found to be $\ang{;;0.173}$ and $\ang{;;0.139}$ respectively with mean value of $\sim\ang{;;0}$ in both the cases. A detailed description and analysis of the astrometric calibration will be presented in Kumar et al. (in prep). 

\subsection{Photometric Calibration catalog}
\label{sec:photometic_calibration_catalog}

Photometric calibration is usually performed by observing the Landolt equatorial standards that have precisely estimated magnitudes \citep{landolt1992}. But these photometric standard stars are mostly located close to the celestial equator and very few of these sources lie inside the ILMT strip. Hence, it is not feasible to calibrate the photometric data of ILMT using this approach.

\begin{figure}[htbp]
    \centering
    \includegraphics{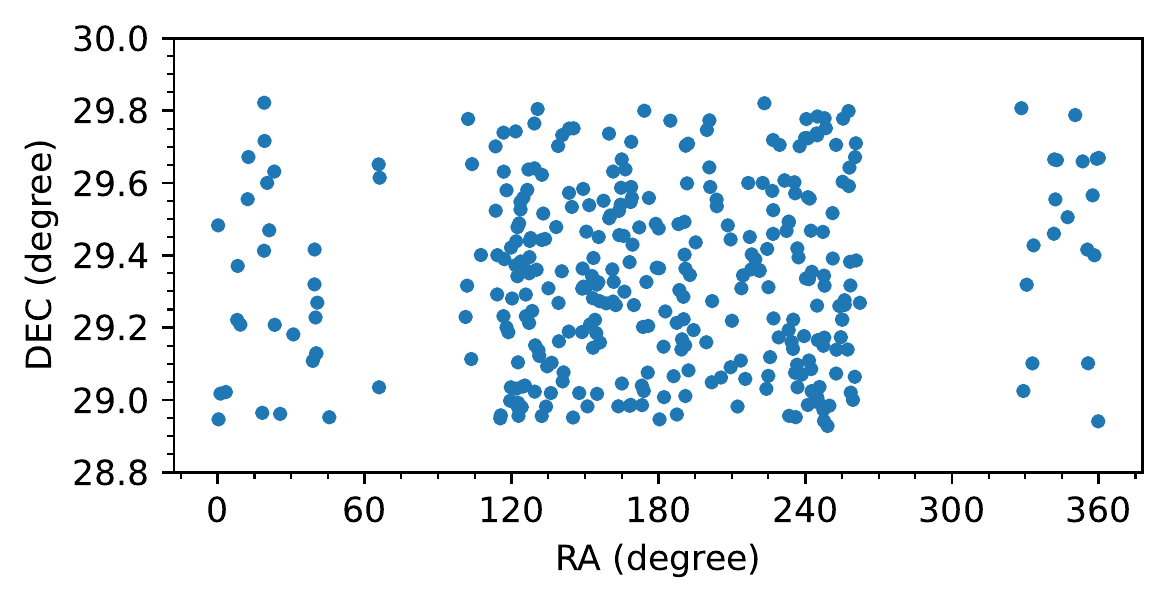}
    \caption{Distribution of spectroscopically confirmed white dwarfs in the ILMT strip.}
    \label{fig:mwdd}
\end{figure}

We, therefore, present a catalog based on white dwarfs that may be used for the photometric calibration of the sources present in the ILMT strip. 
White dwarfs possess a number of properties that make them good calibration standards. These properties include stable photometric magnitudes (except for \emph{DAV} and \emph{DBV} variables), and energy distribution from near-IR to UV. Also, most of them would be fainter than the bright limit of ILMT. We have generated a separate catalog of spectroscopically confirmed white dwarfs by crossmatching our catalog with the Montreal White Dwarf Database (MWDD) \citep{mwdd} which is a collection of spectroscopically confirmed white dwarfs and their properties aggregated from throughout the literature. By doing this, we have the $g, r$, and $i$ photometry of these white dwarfs available and it can be used to calibrate the ILMT fields. There are 345 such white dwarfs in the ILMT strip which are shown in Figure \ref{fig:mwdd}. This catalog can be a part of the photometric data reduction pipeline in due course of time.

\section{Summary}
\label{sec:summary}
The ILMT is a survey telescope that will observe $\sim 120$ square degree strip of sky, centered at declination \ang{+29;21;41}, in three optical bands. In order to facilitate astrometric calibration of the observed images, we have compiled a catalog of astrometric calibrators in this region. This catalog lists precise source positions from \textit{Gaia}, and photometric magnitudes from SDSS or PanSTARRS1 in $g, r$ \& $i$ filters in the ILMT strip. We have achieved this by cross-matching \textit{Gaia} data with secondary catalogs. 

The final catalog contains $\sim 3.56$ million sources most of which are at low galactic latitudes.  All the sources have parallax $< 10$ mas, proper motion $< 20$ mas/yr and $i$ band magnitude between $16.5$ and $22$ $mag$. The data are stored in a local server in SQLite databases for query-based access. The preliminary analysis shows that standard deviations of offsets after performing astrometric calibration using this catalog are around $\sim\ang{;;0.15}$.

Apart from astrometric calibration, this catalog can also be used for photometric calibration. One such method would be to employ white dwarfs as photometric calibrators. To achieve this, we have also prepared a catalog of spectroscopically confirmed white dwarfs with SDSS photometry that potentially can be used for the photometric calibration of the ILMT survey. A complete catalog of all sources that can be used as photometric calibrators will be prepared in the future.

\section*{Acknowledgements}

The 4m International Liquid Mirror Telescope (ILMT) project results from a collaboration between Aryabhatta Research Institute of Observational Sciences (ARIES, India), the Institute of Astrophysics and Geophysics (Li\`{e}ge University), the Canadian Astronomical Institutes, University of Montreal, University of Toronto, York University, University of British Columbia and Victoria University.

PH acknowledges financial support from the Natural Sciences and Engineering Research Council of Canada, RGPIN-2019-04369.

This work has made use of data from the European Space Agency (ESA) mission
{\it Gaia} (\url{https://www.cosmos.esa.int/gaia}), processed by the {\it Gaia}
Data Processing and Analysis Consortium (DPAC, \url{https://www.cosmos.esa.int/web/gaia/dpac/consortium}). Funding for the DPAC has been provided by national institutions, in particular the institutions participating in the {\it Gaia} Multilateral Agreement.

Funding for the Sloan Digital Sky 
Survey IV has been provided by the 
Alfred P. Sloan Foundation, the U.S. 
Department of Energy Office of 
Science, and the Participating 
Institutions. SDSS-IV acknowledges support and  resources from the Center for High Performance Computing  at the University of Utah. The SDSS website is www.sdss.org.

SDSS-IV is managed by the Astrophysical Research Consortium for the Participating Institutions of the SDSS Collaboration including the Brazilian Participation Group, the Carnegie Institution for Science, Carnegie Mellon University, Center for Astrophysics | Harvard \& Smithsonian, the Chilean Participation Group, the French Participation Group, Instituto de Astrof\'isica de Canarias, The Johns Hopkins University, Kavli Institute for the Physics and Mathematics of the Universe (IPMU) / University of Tokyo, the Korean Participation Group, Lawrence Berkeley National Laboratory,  Leibniz Institut f\"ur Astrophysik Potsdam (AIP), Max-Planck-Institut f\"ur Astronomie (MPIA Heidelberg), Max-Planck-Institut f\"ur Astrophysik (MPA Garching), Max-Planck-Institut f\"ur Extraterrestrische Physik (MPE), National Astronomical Observatories of China, New Mexico State University, New York University, University of Notre Dame, Observat\'ario Nacional / MCTI, The Ohio State University, Pennsylvania State University, Shanghai Astronomical Observatory, United Kingdom Participation Group, Universidad Nacional Aut\'onoma de M\'exico, University of Arizona, University of Colorado Boulder, University of Oxford, University of Portsmouth, University of Utah, University of Virginia, University of Washington, University of Wisconsin, Vanderbilt University, and Yale University.

The Pan-STARRS1 Surveys (PS1) and the PS1 public science archive have been made possible through contributions by the Institute for Astronomy, the University of Hawaii, the Pan-STARRS Project Office, the Max-Planck Society and its participating institutes, the Max Planck Institute for Astronomy, Heidelberg and the Max Planck Institute for Extraterrestrial Physics, Garching, The Johns Hopkins University, Durham University, the University of Edinburgh, the Queen's University Belfast, the Harvard-Smithsonian Center for Astrophysics, the Las Cumbres Observatory Global Telescope Network Incorporated, the National Central University of Taiwan, the Space Telescope Science Institute, the National Aeronautics and Space Administration under Grant No. NNX08AR22G issued through the Planetary Science Division of the NASA Science Mission Directorate, the National Science Foundation Grant No. AST-1238877, the University of Maryland, Eotvos Lorand University (ELTE), the Los Alamos National Laboratory, and the Gordon and Betty Moore Foundation.

\bibliographystyle{ws-jai.bst}
\setlength{\bibsep}{0pt plus 0.3ex}
{\footnotesize
\bibliography{bib}

\begin{thebibliography}{24}
\newcommand{\enquote}[1]{``#1''}
\providecommand{\natexlab}[1]{#1}
\providecommand{\url}[1]{\texttt{#1}}
\providecommand{\urlprefix}{URL }
\expandafter\ifx\csname urlstyle\endcsname\relax
  \providecommand{\doi}[1]{doi:\discretionary{}{}{}#1}\else
  \providecommand{\doi}{doi:\discretionary{}{}{}\begingroup
  \urlstyle{rm}\Url}\fi

\bibitem[{{Abdurro'uf} \emph{et~al.}(2022){Abdurro'uf}, {Accetta}, {Aerts},
  {Silva Aguirre}, {Ahumada}, {Ajgaonkar}, {Filiz Ak}, {Alam}, {Allende
  Prieto}, {Almeida}, {Anders}, {Anderson}, {Andrews}, {Anguiano},
  {Aquino-Ort{\'\i}z}, {Arag{\'o}n-Salamanca}, {Argudo-Fern{\'a}ndez}, {Ata},
  {Aubert}, {Avila-Reese}, {Badenes}, {Barb{\'a}}, {Barger},
  {Barrera-Ballesteros}, {Beaton}, {Beers}, {Belfiore}, {Bender}, {Bernardi},
  {Bershady}, {Beutler}, {Bidin}, {Bird}, {Bizyaev}, {Blanc}, {Blanton},
  {Boardman}, {Bolton}, {Boquien}, {Borissova}, {Bovy}, {Brandt}, {Brown},
  {Brownstein}, {Brusa}, {Buchner}, {Bundy}, {Burchett}, {Bureau}, {Burgasser},
  {Cabang}, {Campbell}, {Cappellari}, {Carlberg}, {Wanderley}, {Carrera},
  {Cash}, {Chen}, {Chen}, {Cherinka}, {Chiappini}, {Choi}, {Chojnowski},
  {Chung}, {Clerc}, {Cohen}, {Comerford}, {Comparat}, {da Costa}, {Covey},
  {Crane}, {Cruz-Gonzalez}, {Culhane}, {Cunha}, {Dai}, {Damke}, {Darling},
  {Davidson}, {Davies}, {Dawson}, {De Lee}, {Diamond-Stanic}, {Cano-D{\'\i}az},
  {S{\'a}nchez}, {Donor}, {Duckworth}, {Dwelly}, {Eisenstein}, {Elsworth},
  {Emsellem}, {Eracleous}, {Escoffier}, {Fan}, {Farr}, {Feng},
  {Fern{\'a}ndez-Trincado}, {Feuillet}, {Filipp}, {Fillingham}, {Frinchaboy},
  {Fromenteau}, {Galbany}, {Garc{\'\i}a}, {Garc{\'\i}a-Hern{\'a}ndez}, {Ge},
  {Geisler}, {Gelfand}, {G{\'e}ron}, {Gibson}, {Goddy}, {Godoy-Rivera},
  {Grabowski}, {Green}, {Greener}, {Grier}, {Griffith}, {Guo}, {Guy},
  {Hadjara}, {Harding}, {Hasselquist}, {Hayes}, {Hearty}, {Hern{\'a}ndez},
  {Hill}, {Hogg}, {Holtzman}, {Horta}, {Hsieh}, {Hsu}, {Hsu}, {Huber},
  {Huertas-Company}, {Hutchinson}, {Hwang}, {Ibarra-Medel}, {Chitham}, {Ilha},
  {Imig}, {Jaekle}, {Jayasinghe}, {Ji}, {Johnson}, {Jones}, {J{\"o}nsson},
  {Katkov}, {Khalatyan}, {Kinemuchi}, {Kisku}, {Knapen}, {Kneib}, {Kollmeier},
  {Kong}, {Kounkel}, {Kreckel}, {Krishnarao}, {Lacerna}, {Lane}, {Langgin},
  {Lavender}, {Law}, {Lazarz}, {Leung}, {Leung}, {Lewis}, {Li}, {Li}, {Lian},
  {Liang}, {Lin}, {Lin}, {Lin}, {Lintott}, {Long}, {Longa-Pe{\~n}a},
  {L{\'o}pez-Cob{\'a}}, {Lu}, {Lundgren}, {Luo}, {Mackereth}, {de la Macorra},
  {Mahadevan}, {Majewski}, {Manchado}, {Mandeville}, {Maraston},
  {Margalef-Bentabol}, {Masseron}, {Masters}, {Mathur}, {McDermid}, {Mckay},
  {Merloni}, {Merrifield}, {Meszaros}, {Miglio}, {Di Mille}, {Minniti},
  {Minsley}, {Monachesi}, {Moon}, {Mosser}, {Mulchaey}, {Muna}, {Mu{\~n}oz},
  {Myers}, {Myers}, {Nadathur}, {Nair}, {Nandra}, {Neumann}, {Newman},
  {Nidever}, {Nikakhtar}, {Nitschelm}, {O'Connell}, {Garma-Oehmichen}, {Luan
  Souza de Oliveira}, {Olney}, {Oravetz}, {Ortigoza-Urdaneta}, {Osorio},
  {Otter}, {Pace}, {Padilla}, {Pan}, {Pan}, {Parikh}, {Parker}, {Peirani},
  {Pe{\~n}a Ram{\'\i}rez}, {Penny}, {Percival}, {Perez-Fournon},
  {Pinsonneault}, {Poidevin}, {Poovelil}, {Price-Whelan}, {B{\'a}rbara de
  Andrade Queiroz}, {Raddick}, {Ray}, {Rembold}, {Riddle}, {Riffel}, {Riffel},
  {Rix}, {Robin}, {Rodr{\'\i}guez-Puebla}, {Roman-Lopes},
  {Rom{\'a}n-Z{\'u}{\~n}iga}, {Rose}, {Ross}, {Rossi}, {Rubin}, {Salvato},
  {S{\'a}nchez}, {S{\'a}nchez-Gallego}, {Sanderson}, {Santana Rojas},
  {Sarceno}, {Sarmiento}, {Sayres}, {Sazonova}, {Schaefer}, {Schiavon},
  {Schlegel}, {Schneider}, {Schultheis}, {Schwope}, {Serenelli}, {Serna},
  {Shao}, {Shapiro}, {Sharma}, {Shen}, {Shetrone}, {Shu}, {Simon}, {Skrutskie},
  {Smethurst}, {Smith}, {Sobeck}, {Spoo}, {Sprague}, {Stark}, {Stassun},
  {Steinmetz}, {Stello}, {Stone-Martinez}, {Storchi-Bergmann}, {Stringfellow},
  {Stutz}, {Su}, {Taghizadeh-Popp}, {Talbot}, {Tayar}, {Telles}, {Teske},
  {Thakar}, {Theissen}, {Tkachenko}, {Thomas}, {Tojeiro}, {Hernandez Toledo},
  {Troup}, {Trump}, {Trussler}, {Turner}, {Tuttle}, {Unda-Sanzana},
  {V{\'a}zquez-Mata}, {Valentini}, {Valenzuela}, {Vargas-Gonz{\'a}lez},
  {Vargas-Maga{\~n}a}, {Alfaro}, {Villanova}, {Vincenzo}, {Wake}, {Warfield},
  {Washington}, {Weaver}, {Weijmans}, {Weinberg}, {Weiss}, {Westfall}, {Wild},
  {Wilde}, {Wilson}, {Wilson}, {Wilson}, {Wolf}, {Wood-Vasey}, {Yan}, {Zamora},
  {Zasowski}, {Zhang}, {Zhao}, {Zheng}, {Zheng} \& {Zhu}}]{sdssdr17}
{Abdurro'uf}, {Accetta}, K., {Aerts}, C., {Silva Aguirre}, V., {Ahumada}, R.,
  {Ajgaonkar}, N., {Filiz Ak}, N., {Alam}, S., {Allende Prieto}, C., {Almeida},
  A., {Anders}, F., {Anderson}, S.~F., {Andrews}, B.~H., {Anguiano}, B.,
  {Aquino-Ort{\'\i}z}, E., {Arag{\'o}n-Salamanca}, A., {Argudo-Fern{\'a}ndez},
  M., {Ata}, M., {Aubert}, M., {Avila-Reese}, V., {Badenes}, C., {Barb{\'a}},
  R.~H., {Barger}, K., {Barrera-Ballesteros}, J.~K., {Beaton}, R.~L., {Beers},
  T.~C., {Belfiore}, F., {Bender}, C.~F., {Bernardi}, M., {Bershady}, M.~A.,
  {Beutler}, F., {Bidin}, C.~M., {Bird}, J.~C., {Bizyaev}, D., {Blanc}, G.~A.,
  {Blanton}, M.~R., {Boardman}, N.~F., {Bolton}, A.~S., {Boquien}, M.,
  {Borissova}, J., {Bovy}, J., {Brandt}, W.~N., {Brown}, J., {Brownstein},
  J.~R., {Brusa}, M., {Buchner}, J., {Bundy}, K., {Burchett}, J.~N., {Bureau},
  M., {Burgasser}, A., {Cabang}, T.~K., {Campbell}, S., {Cappellari}, M.,
  {Carlberg}, J.~K., {Wanderley}, F.~C., {Carrera}, R., {Cash}, J., {Chen},
  Y.-P., {Chen}, W.-H., {Cherinka}, B., {Chiappini}, C., {Choi}, P.~D.,
  {Chojnowski}, S.~D., {Chung}, H., {Clerc}, N., {Cohen}, R.~E., {Comerford},
  J.~M., {Comparat}, J., {da Costa}, L., {Covey}, K., {Crane}, J.~D.,
  {Cruz-Gonzalez}, I., {Culhane}, C., {Cunha}, K., {Dai}, Y.~S., {Damke}, G.,
  {Darling}, J., {Davidson}, J., James~W., {Davies}, R., {Dawson}, K., {De
  Lee}, N., {Diamond-Stanic}, A.~M., {Cano-D{\'\i}az}, M., {S{\'a}nchez},
  H.~D., {Donor}, J., {Duckworth}, C., {Dwelly}, T., {Eisenstein}, D.~J.,
  {Elsworth}, Y.~P., {Emsellem}, E., {Eracleous}, M., {Escoffier}, S., {Fan},
  X., {Farr}, E., {Feng}, S., {Fern{\'a}ndez-Trincado}, J.~G., {Feuillet}, D.,
  {Filipp}, A., {Fillingham}, S.~P., {Frinchaboy}, P.~M., {Fromenteau}, S.,
  {Galbany}, L., {Garc{\'\i}a}, R.~A., {Garc{\'\i}a-Hern{\'a}ndez}, D.~A.,
  {Ge}, J., {Geisler}, D., {Gelfand}, J., {G{\'e}ron}, T., {Gibson}, B.~J.,
  {Goddy}, J., {Godoy-Rivera}, D., {Grabowski}, K., {Green}, P.~J., {Greener},
  M., {Grier}, C.~J., {Griffith}, E., {Guo}, H., {Guy}, J., {Hadjara}, M.,
  {Harding}, P., {Hasselquist}, S., {Hayes}, C.~R., {Hearty}, F.,
  {Hern{\'a}ndez}, J., {Hill}, L., {Hogg}, D.~W., {Holtzman}, J.~A., {Horta},
  D., {Hsieh}, B.-C., {Hsu}, C.-H., {Hsu}, Y.-H., {Huber}, D.,
  {Huertas-Company}, M., {Hutchinson}, B., {Hwang}, H.~S., {Ibarra-Medel},
  H.~J., {Chitham}, J.~I., {Ilha}, G.~S., {Imig}, J., {Jaekle}, W.,
  {Jayasinghe}, T., {Ji}, X., {Johnson}, J.~A., {Jones}, A., {J{\"o}nsson}, H.,
  {Katkov}, I., {Khalatyan}, D., Arman, {Kinemuchi}, K., {Kisku}, S., {Knapen},
  J.~H., {Kneib}, J.-P., {Kollmeier}, J.~A., {Kong}, M., {Kounkel}, M.,
  {Kreckel}, K., {Krishnarao}, D., {Lacerna}, I., {Lane}, R.~R., {Langgin}, R.,
  {Lavender}, R., {Law}, D.~R., {Lazarz}, D., {Leung}, H.~W., {Leung}, H.-H.,
  {Lewis}, H.~M., {Li}, C., {Li}, R., {Lian}, J., {Liang}, F.-H., {Lin}, L.,
  {Lin}, Y.-T., {Lin}, S., {Lintott}, C., {Long}, D., {Longa-Pe{\~n}a}, P.,
  {L{\'o}pez-Cob{\'a}}, C., {Lu}, S., {Lundgren}, B.~F., {Luo}, Y.,
  {Mackereth}, J.~T., {de la Macorra}, A., {Mahadevan}, S., {Majewski}, S.~R.,
  {Manchado}, A., {Mandeville}, T., {Maraston}, C., {Margalef-Bentabol}, B.,
  {Masseron}, T., {Masters}, K.~L., {Mathur}, S., {McDermid}, R.~M., {Mckay},
  M., {Merloni}, A., {Merrifield}, M., {Meszaros}, S., {Miglio}, A., {Di
  Mille}, F., {Minniti}, D., {Minsley}, R., {Monachesi}, A., {Moon}, J.,
  {Mosser}, B., {Mulchaey}, J., {Muna}, D., {Mu{\~n}oz}, R.~R., {Myers}, A.~D.,
  {Myers}, N., {Nadathur}, S., {Nair}, P., {Nandra}, K., {Neumann}, J.,
  {Newman}, J.~A., {Nidever}, D.~L., {Nikakhtar}, F., {Nitschelm}, C.,
  {O'Connell}, J.~E., {Garma-Oehmichen}, L., {Luan Souza de Oliveira}, G.,
  {Olney}, R., {Oravetz}, D., {Ortigoza-Urdaneta}, M., {Osorio}, Y., {Otter},
  J., {Pace}, Z.~J., {Padilla}, N., {Pan}, K., {Pan}, H.-A., {Parikh}, T.,
  {Parker}, J., {Peirani}, S., {Pe{\~n}a Ram{\'\i}rez}, K., {Penny}, S.,
  {Percival}, W.~J., {Perez-Fournon}, I., {Pinsonneault}, M., {Poidevin}, F.,
  {Poovelil}, V.~J., {Price-Whelan}, A.~M., {B{\'a}rbara de Andrade Queiroz},
  A., {Raddick}, M.~J., {Ray}, A., {Rembold}, S.~B., {Riddle}, N., {Riffel},
  R.~A., {Riffel}, R., {Rix}, H.-W., {Robin}, A.~C., {Rodr{\'\i}guez-Puebla},
  A., {Roman-Lopes}, A., {Rom{\'a}n-Z{\'u}{\~n}iga}, C., {Rose}, B., {Ross},
  A.~J., {Rossi}, G., {Rubin}, K. H.~R., {Salvato}, M., {S{\'a}nchez}, S.~F.,
  {S{\'a}nchez-Gallego}, J.~R., {Sanderson}, R., {Santana Rojas}, F.~A.,
  {Sarceno}, E., {Sarmiento}, R., {Sayres}, C., {Sazonova}, E., {Schaefer},
  A.~L., {Schiavon}, R., {Schlegel}, D.~J., {Schneider}, D.~P., {Schultheis},
  M., {Schwope}, A., {Serenelli}, A., {Serna}, J., {Shao}, Z., {Shapiro}, G.,
  {Sharma}, A., {Shen}, Y., {Shetrone}, M., {Shu}, Y., {Simon}, J.~D.,
  {Skrutskie}, M.~F., {Smethurst}, R., {Smith}, V., {Sobeck}, J., {Spoo}, T.,
  {Sprague}, D., {Stark}, D.~V., {Stassun}, K.~G., {Steinmetz}, M., {Stello},
  D., {Stone-Martinez}, A., {Storchi-Bergmann}, T., {Stringfellow}, G.~S.,
  {Stutz}, A., {Su}, Y.-C., {Taghizadeh-Popp}, M., {Talbot}, M.~S., {Tayar},
  J., {Telles}, E., {Teske}, J., {Thakar}, A., {Theissen}, C., {Tkachenko}, A.,
  {Thomas}, D., {Tojeiro}, R., {Hernandez Toledo}, H., {Troup}, N.~W., {Trump},
  J.~R., {Trussler}, J., {Turner}, J., {Tuttle}, S., {Unda-Sanzana}, E.,
  {V{\'a}zquez-Mata}, J.~A., {Valentini}, M., {Valenzuela}, O.,
  {Vargas-Gonz{\'a}lez}, J., {Vargas-Maga{\~n}a}, M., {Alfaro}, P.~V.,
  {Villanova}, S., {Vincenzo}, F., {Wake}, D., {Warfield}, J.~T., {Washington},
  J.~D., {Weaver}, B.~A., {Weijmans}, A.-M., {Weinberg}, D.~H., {Weiss}, A.,
  {Westfall}, K.~B., {Wild}, V., {Wilde}, M.~C., {Wilson}, J.~C., {Wilson},
  R.~F., {Wilson}, M., {Wolf}, J., {Wood-Vasey}, W.~M., {Yan}, R., {Zamora},
  O., {Zasowski}, G., {Zhang}, K., {Zhao}, C., {Zheng}, Z., {Zheng}, Z. \&
  {Zhu}, K. [2022]  \emph{\apjs} \textbf{259}, 35,
  \doi{10.3847/1538-4365/ac4414}.

\bibitem[{{Blanton} \emph{et~al.}(2017){Blanton}, {Bershady}, {Abolfathi},
  {Albareti}, {Allende Prieto}, {Almeida}, {Alonso-Garc{\'\i}a}, {Anders},
  {Anderson}, {Andrews}, {Aquino-Ort{\'\i}z}, {Arag{\'o}n-Salamanca},
  {Argudo-Fern{\'a}ndez}, {Armengaud}, {Aubourg}, {Avila-Reese}, {Badenes},
  {Bailey}, {Barger}, {Barrera-Ballesteros}, {Bartosz}, {Bates}, {Baumgarten},
  {Bautista}, {Beaton}, {Beers}, {Belfiore}, {Bender}, {Berlind}, {Bernardi},
  {Beutler}, {Bird}, {Bizyaev}, {Blanc}, {Blomqvist}, {Bolton}, {Boquien},
  {Borissova}, {van den Bosch}, {Bovy}, {Brandt}, {Brinkmann}, {Brownstein},
  {Bundy}, {Burgasser}, {Burtin}, {Busca}, {Cappellari}, {Delgado Carigi},
  {Carlberg}, {Carnero Rosell}, {Carrera}, {Chanover}, {Cherinka}, {Cheung},
  {G{\'o}mez Maqueo Chew}, {Chiappini}, {Choi}, {Chojnowski}, {Chuang},
  {Chung}, {Cirolini}, {Clerc}, {Cohen}, {Comparat}, {da Costa}, {Cousinou},
  {Covey}, {Crane}, {Croft}, {Cruz-Gonzalez}, {Garrido Cuadra}, {Cunha},
  {Damke}, {Darling}, {Davies}, {Dawson}, {de la Macorra}, {Dell'Agli}, {De
  Lee}, {Delubac}, {Di Mille}, {Diamond-Stanic}, {Cano-D{\'\i}az}, {Donor},
  {Downes}, {Drory}, {du Mas des Bourboux}, {Duckworth}, {Dwelly}, {Dyer},
  {Ebelke}, {Eigenbrot}, {Eisenstein}, {Emsellem}, {Eracleous}, {Escoffier},
  {Evans}, {Fan}, {Fern{\'a}ndez-Alvar}, {Fernandez-Trincado}, {Feuillet},
  {Finoguenov}, {Fleming}, {Font-Ribera}, {Fredrickson}, {Freischlad},
  {Frinchaboy}, {Fuentes}, {Galbany}, {Garcia-Dias},
  {Garc{\'\i}a-Hern{\'a}ndez}, {Gaulme}, {Geisler}, {Gelfand},
  {Gil-Mar{\'\i}n}, {Gillespie}, {Goddard}, {Gonzalez-Perez}, {Grabowski},
  {Green}, {Grier}, {Gunn}, {Guo}, {Guy}, {Hagen}, {Hahn}, {Hall}, {Harding},
  {Hasselquist}, {Hawley}, {Hearty}, {Gonzalez Hern{\'a}ndez}, {Ho}, {Hogg},
  {Holley-Bockelmann}, {Holtzman}, {Holzer}, {Huehnerhoff}, {Hutchinson},
  {Hwang}, {Ibarra-Medel}, {da Silva Ilha}, {Ivans}, {Ivory}, {Jackson},
  {Jensen}, {Johnson}, {Jones}, {J{\"o}nsson}, {Jullo}, {Kamble}, {Kinemuchi},
  {Kirkby}, {Kitaura}, {Klaene}, {Knapp}, {Kneib}, {Kollmeier}, {Lacerna},
  {Lane}, {Lang}, {Law}, {Lazarz}, {Lee}, {Le Goff}, {Liang}, {Li}, {Li},
  {Lian}, {Lima}, {Lin}, {Lin}, {Bertran de Lis}, {Liu}, {de Icaza Lizaola},
  {Long}, {Lucatello}, {Lundgren}, {MacDonald}, {Deconto Machado}, {MacLeod},
  {Mahadevan}, {Geimba Maia}, {Maiolino}, {Majewski}, {Malanushenko},
  {Malanushenko}, {Manchado}, {Mao}, {Maraston}, {Marques-Chaves}, {Masseron},
  {Masters}, {McBride}, {McDermid}, {McGrath}, {McGreer}, {Medina Pe{\~n}a},
  {Melendez}, {Merloni}, {Merrifield}, {Meszaros}, {Meza}, {Minchev},
  {Minniti}, {Miyaji}, {More}, {Mulchaey}, {M{\"u}ller-S{\'a}nchez}, {Muna},
  {Munoz}, {Myers}, {Nair}, {Nandra}, {Correa do Nascimento}, {Negrete},
  {Ness}, {Newman}, {Nichol}, {Nidever}, {Nitschelm}, {Ntelis}, {O'Connell},
  {Oelkers}, {Oravetz}, {Oravetz}, {Pace}, {Padilla}, {Palanque-Delabrouille},
  {Alonso Palicio}, {Pan}, {Parejko}, {Parikh}, {P{\^a}ris}, {Park}, {Patten},
  {Peirani}, {Pellejero-Ibanez}, {Penny}, {Percival}, {Perez-Fournon},
  {Petitjean}, {Pieri}, {Pinsonneault}, {Pisani}, {Poleski}, {Prada},
  {Prakash}, {Queiroz}, {Raddick}, {Raichoor}, {Barboza Rembold}, {Richstein},
  {Riffel}, {Riffel}, {Rix}, {Robin}, {Rockosi}, {Rodr{\'\i}guez-Torres},
  {Roman-Lopes}, {Rom{\'a}n-Z{\'u}{\~n}iga}, {Rosado}, {Ross}, {Rossi}, {Ruan},
  {Ruggeri}, {Rykoff}, {Salazar-Albornoz}, {Salvato}, {S{\'a}nchez}, {Aguado},
  {S{\'a}nchez-Gallego}, {Santana}, {Santiago}, {Sayres}, {Schiavon}, {da Silva
  Schimoia}, {Schlafly}, {Schlegel}, {Schneider}, {Schultheis}, {Schuster},
  {Schwope}, {Seo}, {Shao}, {Shen}, {Shetrone}, {Shull}, {Simon}, {Skinner},
  {Skrutskie}, {Slosar}, {Smith}, {Sobeck}, {Sobreira}, {Somers}, {Souto},
  {Stark}, {Stassun}, {Stauffer}, {Steinmetz}, {Storchi-Bergmann},
  {Streblyanska}, {Stringfellow}, {Su{\'a}rez}, {Sun}, {Suzuki}, {Szigeti},
  {Taghizadeh-Popp}, {Tang}, {Tao}, {Tayar}, {Tembe}, {Teske}, {Thakar},
  {Thomas}, {Thompson}, {Tinker}, {Tissera}, {Tojeiro}, {Hernandez Toledo}, {de
  la Torre}, {Tremonti}, {Troup}, {Valenzuela}, {Martinez Valpuesta},
  {Vargas-Gonz{\'a}lez}, {Vargas-Maga{\~n}a}, {Vazquez}, {Villanova}, {Vivek},
  {Vogt}, {Wake}, {Walterbos}, {Wang}, {Weaver}, {Weijmans}, {Weinberg},
  {Westfall}, {Whelan}, {Wild}, {Wilson}, {Wood-Vasey}, {Wylezalek}, {Xiao},
  {Yan}, {Yang}, {Ybarra}, {Y{\`e}che}, {Zakamska}, {Zamora}, {Zarrouk},
  {Zasowski}, {Zhang}, {Zhao}, {Zheng}, {Zheng}, {Zhou}, {Zhou}, {Zhu},
  {Zoccali} \& {Zou}}]{sdssIV}
{Blanton}, M.~R., {Bershady}, M.~A., {Abolfathi}, B., {Albareti}, F.~D.,
  {Allende Prieto}, C., {Almeida}, A., {Alonso-Garc{\'\i}a}, J., {Anders}, F.,
  {Anderson}, S.~F., {Andrews}, B., {Aquino-Ort{\'\i}z}, E.,
  {Arag{\'o}n-Salamanca}, A., {Argudo-Fern{\'a}ndez}, M., {Armengaud}, E.,
  {Aubourg}, E., {Avila-Reese}, V., {Badenes}, C., {Bailey}, S., {Barger},
  K.~A., {Barrera-Ballesteros}, J., {Bartosz}, C., {Bates}, D., {Baumgarten},
  F., {Bautista}, J., {Beaton}, R., {Beers}, T.~C., {Belfiore}, F., {Bender},
  C.~F., {Berlind}, A.~A., {Bernardi}, M., {Beutler}, F., {Bird}, J.~C.,
  {Bizyaev}, D., {Blanc}, G.~A., {Blomqvist}, M., {Bolton}, A.~S., {Boquien},
  M., {Borissova}, J., {van den Bosch}, R., {Bovy}, J., {Brandt}, W.~N.,
  {Brinkmann}, J., {Brownstein}, J.~R., {Bundy}, K., {Burgasser}, A.~J.,
  {Burtin}, E., {Busca}, N.~G., {Cappellari}, M., {Delgado Carigi}, M.~L.,
  {Carlberg}, J.~K., {Carnero Rosell}, A., {Carrera}, R., {Chanover}, N.~J.,
  {Cherinka}, B., {Cheung}, E., {G{\'o}mez Maqueo Chew}, Y., {Chiappini}, C.,
  {Choi}, P.~D., {Chojnowski}, D., {Chuang}, C.-H., {Chung}, H., {Cirolini},
  R.~F., {Clerc}, N., {Cohen}, R.~E., {Comparat}, J., {da Costa}, L.,
  {Cousinou}, M.-C., {Covey}, K., {Crane}, J.~D., {Croft}, R. A.~C.,
  {Cruz-Gonzalez}, I., {Garrido Cuadra}, D., {Cunha}, K., {Damke}, G.~J.,
  {Darling}, J., {Davies}, R., {Dawson}, K., {de la Macorra}, A., {Dell'Agli},
  F., {De Lee}, N., {Delubac}, T., {Di Mille}, F., {Diamond-Stanic}, A.,
  {Cano-D{\'\i}az}, M., {Donor}, J., {Downes}, J.~J., {Drory}, N., {du Mas des
  Bourboux}, H., {Duckworth}, C.~J., {Dwelly}, T., {Dyer}, J., {Ebelke}, G.,
  {Eigenbrot}, A.~D., {Eisenstein}, D.~J., {Emsellem}, E., {Eracleous}, M.,
  {Escoffier}, S., {Evans}, M.~L., {Fan}, X., {Fern{\'a}ndez-Alvar}, E.,
  {Fernandez-Trincado}, J.~G., {Feuillet}, D.~K., {Finoguenov}, A., {Fleming},
  S.~W., {Font-Ribera}, A., {Fredrickson}, A., {Freischlad}, G., {Frinchaboy},
  P.~M., {Fuentes}, C.~E., {Galbany}, L., {Garcia-Dias}, R.,
  {Garc{\'\i}a-Hern{\'a}ndez}, D.~A., {Gaulme}, P., {Geisler}, D., {Gelfand},
  J.~D., {Gil-Mar{\'\i}n}, H., {Gillespie}, B.~A., {Goddard}, D.,
  {Gonzalez-Perez}, V., {Grabowski}, K., {Green}, P.~J., {Grier}, C.~J.,
  {Gunn}, J.~E., {Guo}, H., {Guy}, J., {Hagen}, A., {Hahn}, C., {Hall}, M.,
  {Harding}, P., {Hasselquist}, S., {Hawley}, S.~L., {Hearty}, F., {Gonzalez
  Hern{\'a}ndez}, J.~I., {Ho}, S., {Hogg}, D.~W., {Holley-Bockelmann}, K.,
  {Holtzman}, J.~A., {Holzer}, P.~H., {Huehnerhoff}, J., {Hutchinson}, T.~A.,
  {Hwang}, H.~S., {Ibarra-Medel}, H.~J., {da Silva Ilha}, G., {Ivans}, I.~I.,
  {Ivory}, K., {Jackson}, K., {Jensen}, T.~W., {Johnson}, J.~A., {Jones}, A.,
  {J{\"o}nsson}, H., {Jullo}, E., {Kamble}, V., {Kinemuchi}, K., {Kirkby}, D.,
  {Kitaura}, F.-S., {Klaene}, M., {Knapp}, G.~R., {Kneib}, J.-P., {Kollmeier},
  J.~A., {Lacerna}, I., {Lane}, R.~R., {Lang}, D., {Law}, D.~R., {Lazarz}, D.,
  {Lee}, Y., {Le Goff}, J.-M., {Liang}, F.-H., {Li}, C., {Li}, H., {Lian}, J.,
  {Lima}, M., {Lin}, L., {Lin}, Y.-T., {Bertran de Lis}, S., {Liu}, C., {de
  Icaza Lizaola}, M. A.~C., {Long}, D., {Lucatello}, S., {Lundgren}, B.,
  {MacDonald}, N.~K., {Deconto Machado}, A., {MacLeod}, C.~L., {Mahadevan}, S.,
  {Geimba Maia}, M.~A., {Maiolino}, R., {Majewski}, S.~R., {Malanushenko}, E.,
  {Malanushenko}, V., {Manchado}, A., {Mao}, S., {Maraston}, C.,
  {Marques-Chaves}, R., {Masseron}, T., {Masters}, K.~L., {McBride}, C.~K.,
  {McDermid}, R.~M., {McGrath}, B., {McGreer}, I.~D., {Medina Pe{\~n}a}, N.,
  {Melendez}, M., {Merloni}, A., {Merrifield}, M.~R., {Meszaros}, S., {Meza},
  A., {Minchev}, I., {Minniti}, D., {Miyaji}, T., {More}, S., {Mulchaey}, J.,
  {M{\"u}ller-S{\'a}nchez}, F., {Muna}, D., {Munoz}, R.~R., {Myers}, A.~D.,
  {Nair}, P., {Nandra}, K., {Correa do Nascimento}, J., {Negrete}, A., {Ness},
  M., {Newman}, J.~A., {Nichol}, R.~C., {Nidever}, D.~L., {Nitschelm}, C.,
  {Ntelis}, P., {O'Connell}, J.~E., {Oelkers}, R.~J., {Oravetz}, A., {Oravetz},
  D., {Pace}, Z., {Padilla}, N., {Palanque-Delabrouille}, N., {Alonso Palicio},
  P., {Pan}, K., {Parejko}, J.~K., {Parikh}, T., {P{\^a}ris}, I., {Park}, C.,
  {Patten}, A.~Y., {Peirani}, S., {Pellejero-Ibanez}, M., {Penny}, S.,
  {Percival}, W.~J., {Perez-Fournon}, I., {Petitjean}, P., {Pieri}, M.~M.,
  {Pinsonneault}, M., {Pisani}, A., {Poleski}, R., {Prada}, F., {Prakash}, A.,
  {Queiroz}, A. B. d.~A., {Raddick}, M.~J., {Raichoor}, A., {Barboza Rembold},
  S., {Richstein}, H., {Riffel}, R.~A., {Riffel}, R., {Rix}, H.-W., {Robin},
  A.~C., {Rockosi}, C.~M., {Rodr{\'\i}guez-Torres}, S., {Roman-Lopes}, A.,
  {Rom{\'a}n-Z{\'u}{\~n}iga}, C., {Rosado}, M., {Ross}, A.~J., {Rossi}, G.,
  {Ruan}, J., {Ruggeri}, R., {Rykoff}, E.~S., {Salazar-Albornoz}, S.,
  {Salvato}, M., {S{\'a}nchez}, A.~G., {Aguado}, D.~S., {S{\'a}nchez-Gallego},
  J.~R., {Santana}, F.~A., {Santiago}, B.~X., {Sayres}, C., {Schiavon}, R.~P.,
  {da Silva Schimoia}, J., {Schlafly}, E.~F., {Schlegel}, D.~J., {Schneider},
  D.~P., {Schultheis}, M., {Schuster}, W.~J., {Schwope}, A., {Seo}, H.-J.,
  {Shao}, Z., {Shen}, S., {Shetrone}, M., {Shull}, M., {Simon}, J.~D.,
  {Skinner}, D., {Skrutskie}, M.~F., {Slosar}, A., {Smith}, V.~V., {Sobeck},
  J.~S., {Sobreira}, F., {Somers}, G., {Souto}, D., {Stark}, D.~V., {Stassun},
  K., {Stauffer}, F., {Steinmetz}, M., {Storchi-Bergmann}, T., {Streblyanska},
  A., {Stringfellow}, G.~S., {Su{\'a}rez}, G., {Sun}, J., {Suzuki}, N.,
  {Szigeti}, L., {Taghizadeh-Popp}, M., {Tang}, B., {Tao}, C., {Tayar}, J.,
  {Tembe}, M., {Teske}, J., {Thakar}, A.~R., {Thomas}, D., {Thompson}, B.~A.,
  {Tinker}, J.~L., {Tissera}, P., {Tojeiro}, R., {Hernandez Toledo}, H., {de la
  Torre}, S., {Tremonti}, C., {Troup}, N.~W., {Valenzuela}, O., {Martinez
  Valpuesta}, I., {Vargas-Gonz{\'a}lez}, J., {Vargas-Maga{\~n}a}, M.,
  {Vazquez}, J.~A., {Villanova}, S., {Vivek}, M., {Vogt}, N., {Wake}, D.,
  {Walterbos}, R., {Wang}, Y., {Weaver}, B.~A., {Weijmans}, A.-M., {Weinberg},
  D.~H., {Westfall}, K.~B., {Whelan}, D.~G., {Wild}, V., {Wilson}, J.,
  {Wood-Vasey}, W.~M., {Wylezalek}, D., {Xiao}, T., {Yan}, R., {Yang}, M.,
  {Ybarra}, J.~E., {Y{\`e}che}, C., {Zakamska}, N., {Zamora}, O., {Zarrouk},
  P., {Zasowski}, G., {Zhang}, K., {Zhao}, G.-B., {Zheng}, Z., {Zheng}, Z.,
  {Zhou}, X., {Zhou}, Z.-M., {Zhu}, G.~B., {Zoccali}, M. \& {Zou}, H. [2017]
  \emph{\aj} \textbf{154}, 28, \doi{10.3847/1538-3881/aa7567}.

\bibitem[{{Chambers} \emph{et~al.}(2016){Chambers}, {Magnier}, {Metcalfe},
  {Flewelling}, {Huber}, {Waters}, {Denneau}, {Draper}, {Farrow}, {Finkbeiner},
  {Holmberg}, {Koppenhoefer}, {Price}, {Rest}, {Saglia}, {Schlafly}, {Smartt},
  {Sweeney}, {Wainscoat}, {Burgett}, {Chastel}, {Grav}, {Heasley}, {Hodapp},
  {Jedicke}, {Kaiser}, {Kudritzki}, {Luppino}, {Lupton}, {Monet}, {Morgan},
  {Onaka}, {Shiao}, {Stubbs}, {Tonry}, {White}, {Ba{\~n}ados}, {Bell},
  {Bender}, {Bernard}, {Boegner}, {Boffi}, {Botticella}, {Calamida},
  {Casertano}, {Chen}, {Chen}, {Cole}, {Deacon}, {Frenk}, {Fitzsimmons},
  {Gezari}, {Gibbs}, {Goessl}, {Goggia}, {Gourgue}, {Goldman}, {Grant},
  {Grebel}, {Hambly}, {Hasinger}, {Heavens}, {Heckman}, {Henderson}, {Henning},
  {Holman}, {Hopp}, {Ip}, {Isani}, {Jackson}, {Keyes}, {Koekemoer}, {Kotak},
  {Le}, {Liska}, {Long}, {Lucey}, {Liu}, {Martin}, {Masci}, {McLean}, {Mindel},
  {Misra}, {Morganson}, {Murphy}, {Obaika}, {Narayan}, {Nieto-Santisteban},
  {Norberg}, {Peacock}, {Pier}, {Postman}, {Primak}, {Rae}, {Rai}, {Riess},
  {Riffeser}, {Rix}, {R{\"o}ser}, {Russel}, {Rutz}, {Schilbach}, {Schultz},
  {Scolnic}, {Strolger}, {Szalay}, {Seitz}, {Small}, {Smith}, {Soderblom},
  {Taylor}, {Thomson}, {Taylor}, {Thakar}, {Thiel}, {Thilker}, {Unger},
  {Urata}, {Valenti}, {Wagner}, {Walder}, {Walter}, {Watters}, {Werner},
  {Wood-Vasey} \& {Wyse}}]{panstarrs_survey}
{Chambers}, K.~C., {Magnier}, E.~A., {Metcalfe}, N., {Flewelling}, H.~A.,
  {Huber}, M.~E., {Waters}, C.~Z., {Denneau}, L., {Draper}, P.~W., {Farrow},
  D., {Finkbeiner}, D.~P., {Holmberg}, C., {Koppenhoefer}, J., {Price}, P.~A.,
  {Rest}, A., {Saglia}, R.~P., {Schlafly}, E.~F., {Smartt}, S.~J., {Sweeney},
  W., {Wainscoat}, R.~J., {Burgett}, W.~S., {Chastel}, S., {Grav}, T.,
  {Heasley}, J.~N., {Hodapp}, K.~W., {Jedicke}, R., {Kaiser}, N., {Kudritzki},
  R.~P., {Luppino}, G.~A., {Lupton}, R.~H., {Monet}, D.~G., {Morgan}, J.~S.,
  {Onaka}, P.~M., {Shiao}, B., {Stubbs}, C.~W., {Tonry}, J.~L., {White}, R.,
  {Ba{\~n}ados}, E., {Bell}, E.~F., {Bender}, R., {Bernard}, E.~J., {Boegner},
  M., {Boffi}, F., {Botticella}, M.~T., {Calamida}, A., {Casertano}, S.,
  {Chen}, W.~P., {Chen}, X., {Cole}, S., {Deacon}, N., {Frenk}, C.,
  {Fitzsimmons}, A., {Gezari}, S., {Gibbs}, V., {Goessl}, C., {Goggia}, T.,
  {Gourgue}, R., {Goldman}, B., {Grant}, P., {Grebel}, E.~K., {Hambly}, N.~C.,
  {Hasinger}, G., {Heavens}, A.~F., {Heckman}, T.~M., {Henderson}, R.,
  {Henning}, T., {Holman}, M., {Hopp}, U., {Ip}, W.~H., {Isani}, S., {Jackson},
  M., {Keyes}, C.~D., {Koekemoer}, A.~M., {Kotak}, R., {Le}, D., {Liska}, D.,
  {Long}, K.~S., {Lucey}, J.~R., {Liu}, M., {Martin}, N.~F., {Masci}, G.,
  {McLean}, B., {Mindel}, E., {Misra}, P., {Morganson}, E., {Murphy}, D.~N.~A.,
  {Obaika}, A., {Narayan}, G., {Nieto-Santisteban}, M.~A., {Norberg}, P.,
  {Peacock}, J.~A., {Pier}, E.~A., {Postman}, M., {Primak}, N., {Rae}, C.,
  {Rai}, A., {Riess}, A., {Riffeser}, A., {Rix}, H.~W., {R{\"o}ser}, S.,
  {Russel}, R., {Rutz}, L., {Schilbach}, E., {Schultz}, A.~S.~B., {Scolnic},
  D., {Strolger}, L., {Szalay}, A., {Seitz}, S., {Small}, E., {Smith}, K.~W.,
  {Soderblom}, D.~R., {Taylor}, P., {Thomson}, R., {Taylor}, A.~N., {Thakar},
  A.~R., {Thiel}, J., {Thilker}, D., {Unger}, D., {Urata}, Y., {Valenti}, J.,
  {Wagner}, J., {Walder}, T., {Walter}, F., {Watters}, S.~P., {Werner}, S.,
  {Wood-Vasey}, W.~M. \& {Wyse}, R. [2016]  \emph{arXiv e-prints} ,
  arXiv:1612.05560.

\bibitem[{Dufour \emph{et~al.}(2016)Dufour, Blouin, Coutu, Fortin-Archambault,
  Thibeault, Bergeron \& Fontaine}]{mwdd}
Dufour, P., Blouin, S., Coutu, S., Fortin-Archambault, M., Thibeault, C.,
  Bergeron, P. \& Fontaine, G. [2016]  \enquote{The montreal white dwarf
  database: a tool for the community,} \doi{10.48550/ARXIV.1610.00986},
  \urlprefix\url{https://arxiv.org/abs/1610.00986}.

\bibitem[{{ESA}()}]{1997ESA}
{ESA} [1997]  \emph{{The HIPPARCOS and TYCHO catalogues. Astrometric and
  photometric star catalogues derived from the ESA HIPPARCOS Space Astrometry
  Mission}}, ESA Special Publication, Vol.~1200.

\bibitem[{Flesch(2021)}]{flesch2021}
Flesch, E.~W. [2021]  \enquote{The million quasars (milliquas) v7.2 catalogue,
  now with vlass associations. the inclusion of sdss-dr16q quasars is
  detailed,} \doi{10.48550/ARXIV.2105.12985},
  \urlprefix\url{https://arxiv.org/abs/2105.12985}.

\bibitem[{{Flewelling} \emph{et~al.}(2020){Flewelling}, {Magnier}, {Chambers},
  {Heasley}, {Holmberg}, {Huber}, {Sweeney}, {Waters}, {Calamida}, {Casertano},
  {Chen}, {Farrow}, {Hasinger}, {Henderson}, {Long}, {Metcalfe}, {Narayan},
  {Nieto-Santisteban}, {Norberg}, {Rest}, {Saglia}, {Szalay}, {Thakar},
  {Tonry}, {Valenti}, {Werner}, {White}, {Denneau}, {Draper}, {Hodapp},
  {Jedicke}, {Kaiser}, {Kudritzki}, {Price}, {Wainscoat}, {Chastel}, {McLean},
  {Postman} \& {Shiao}}]{panstarrs_data_products}
{Flewelling}, H.~A., {Magnier}, E.~A., {Chambers}, K.~C., {Heasley}, J.~N.,
  {Holmberg}, C., {Huber}, M.~E., {Sweeney}, W., {Waters}, C.~Z., {Calamida},
  A., {Casertano}, S., {Chen}, X., {Farrow}, D., {Hasinger}, G., {Henderson},
  R., {Long}, K.~S., {Metcalfe}, N., {Narayan}, G., {Nieto-Santisteban}, M.~A.,
  {Norberg}, P., {Rest}, A., {Saglia}, R.~P., {Szalay}, A., {Thakar}, A.~R.,
  {Tonry}, J.~L., {Valenti}, J., {Werner}, S., {White}, R., {Denneau}, L.,
  {Draper}, P.~W., {Hodapp}, K.~W., {Jedicke}, R., {Kaiser}, N., {Kudritzki},
  R.~P., {Price}, P.~A., {Wainscoat}, R.~J., {Chastel}, S., {McLean}, B.,
  {Postman}, M. \& {Shiao}, B. [2020]  \emph{\apjs} \textbf{251}, 7,
  \doi{10.3847/1538-4365/abb82d}.

\bibitem[{{Fukugita} \emph{et~al.}(1996){Fukugita}, {Ichikawa}, {Gunn}, {Doi},
  {Shimasaku} \& {Schneider}}]{sdss_photometric_system}
{Fukugita}, M., {Ichikawa}, T., {Gunn}, J.~E., {Doi}, M., {Shimasaku}, K. \&
  {Schneider}, D.~P. [1996]  \emph{\aj} \textbf{111},  1748,
  \doi{10.1086/117915}.

\bibitem[{{Gaia Collaboration} \emph{et~al.}(2021){Gaia Collaboration},
  {Brown}, {Vallenari}, {Prusti}, {de Bruijne}, {Babusiaux}, {Biermann},
  {Creevey}, {Evans}, {Eyer}, {Hutton}, {Jansen}, {Jordi}, {Klioner},
  {Lammers}, {Lindegren}, {Luri}, {Mignard}, {Panem}, {Pourbaix}, {Randich},
  {Sartoretti}, {Soubiran}, {Walton}, {Arenou}, {Bailer-Jones}, {Bastian},
  {Cropper}, {Drimmel}, {Katz}, {Lattanzi}, {van Leeuwen}, {Bakker},
  {Cacciari}, {Casta{\~n}eda}, {De Angeli}, {Ducourant}, {Fabricius},
  {Fouesneau}, {Fr{\'e}mat}, {Guerra}, {Guerrier}, {Guiraud}, {Jean-Antoine
  Piccolo}, {Masana}, {Messineo}, {Mowlavi}, {Nicolas}, {Nienartowicz},
  {Pailler}, {Panuzzo}, {Riclet}, {Roux}, {Seabroke}, {Sordo}, {Tanga},
  {Th{\'e}venin}, {Gracia-Abril}, {Portell}, {Teyssier}, {Altmann}, {Andrae},
  {Bellas-Velidis}, {Benson}, {Berthier}, {Blomme}, {Brugaletta}, {Burgess},
  {Busso}, {Carry}, {Cellino}, {Cheek}, {Clementini}, {Damerdji}, {Davidson},
  {Delchambre}, {Dell'Oro}, {Fern{\'a}ndez-Hern{\'a}ndez}, {Galluccio},
  {Garc{\'\i}a-Lario}, {Garcia-Reinaldos}, {Gonz{\'a}lez-N{\'u}{\~n}ez},
  {Gosset}, {Haigron}, {Halbwachs}, {Hambly}, {Harrison}, {Hatzidimitriou},
  {Heiter}, {Hern{\'a}ndez}, {Hestroffer}, {Hodgkin}, {Holl}, {Jan{\ss}en},
  {Jevardat de Fombelle}, {Jordan}, {Krone-Martins}, {Lanzafame},
  {L{\"o}ffler}, {Lorca}, {Manteiga}, {Marchal}, {Marrese}, {Moitinho}, {Mora},
  {Muinonen}, {Osborne}, {Pancino}, {Pauwels}, {Petit}, {Recio-Blanco},
  {Richards}, {Riello}, {Rimoldini}, {Robin}, {Roegiers}, {Rybizki}, {Sarro},
  {Siopis}, {Smith}, {Sozzetti}, {Ulla}, {Utrilla}, {van Leeuwen}, {van
  Reeven}, {Abbas}, {Abreu Aramburu}, {Accart}, {Aerts}, {Aguado}, {Ajaj},
  {Altavilla}, {{\'A}lvarez}, {{\'A}lvarez Cid-Fuentes}, {Alves}, {Anderson},
  {Anglada Varela}, {Antoja}, {Audard}, {Baines}, {Baker},
  {Balaguer-N{\'u}{\~n}ez}, {Balbinot}, {Balog}, {Barache}, {Barbato},
  {Barros}, {Barstow}, {Bartolom{\'e}}, {Bassilana}, {Bauchet},
  {Baudesson-Stella}, {Becciani}, {Bellazzini}, {Bernet}, {Bertone}, {Bianchi},
  {Blanco-Cuaresma}, {Boch}, {Bombrun}, {Bossini}, {Bouquillon}, {Bragaglia},
  {Bramante}, {Breedt}, {Bressan}, {Brouillet}, {Bucciarelli}, {Burlacu},
  {Busonero}, {Butkevich}, {Buzzi}, {Caffau}, {Cancelliere}, {C{\'a}novas},
  {Cantat-Gaudin}, {Carballo}, {Carlucci}, {Carnerero}, {Carrasco},
  {Casamiquela}, {Castellani}, {Castro-Ginard}, {Castro Sampol}, {Chaoul},
  {Charlot}, {Chemin}, {Chiavassa}, {Cioni}, {Comoretto}, {Cooper}, {Cornez},
  {Cowell}, {Crifo}, {Crosta}, {Crowley}, {Dafonte}, {Dapergolas}, {David},
  {David}, {de Laverny}, {De Luise}, {De March}, {De Ridder}, {de Souza}, {de
  Teodoro}, {de Torres}, {del Peloso}, {del Pozo}, {Delbo}, {Delgado},
  {Delgado}, {Delisle}, {Di Matteo}, {Diakite}, {Diener}, {Distefano},
  {Dolding}, {Eappachen}, {Edvardsson}, {Enke}, {Esquej}, {Fabre}, {Fabrizio},
  {Faigler}, {Fedorets}, {Fernique}, {Fienga}, {Figueras}, {Fouron},
  {Fragkoudi}, {Fraile}, {Franke}, {Gai}, {Garabato}, {Garcia-Gutierrez},
  {Garc{\'\i}a-Torres}, {Garofalo}, {Gavras}, {Gerlach}, {Geyer}, {Giacobbe},
  {Gilmore}, {Girona}, {Giuffrida}, {Gomel}, {Gomez}, {Gonzalez-Santamaria},
  {Gonz{\'a}lez-Vidal}, {Granvik}, {Guti{\'e}rrez-S{\'a}nchez}, {Guy},
  {Hauser}, {Haywood}, {Helmi}, {Hidalgo}, {Hilger}, {H{\l}adczuk}, {Hobbs},
  {Holland}, {Huckle}, {Jasniewicz}, {Jonker}, {Juaristi Campillo}, {Julbe},
  {Karbevska}, {Kervella}, {Khanna}, {Kochoska}, {Kontizas}, {Kordopatis},
  {Korn}, {Kostrzewa-Rutkowska}, {Kruszy{\'n}ska}, {Lambert}, {Lanza}, {Lasne},
  {Le Campion}, {Le Fustec}, {Lebreton}, {Lebzelter}, {Leccia}, {Leclerc},
  {Lecoeur-Taibi}, {Liao}, {Licata}, {Lindstr{\o}m}, {Lister}, {Livanou},
  {Lobel}, {Madrero Pardo}, {Managau}, {Mann}, {Marchant}, {Marconi}, {Marcos
  Santos}, {Marinoni}, {Marocco}, {Marshall}, {Martin Polo},
  {Mart{\'\i}n-Fleitas}, {Masip}, {Massari}, {Mastrobuono-Battisti}, {Mazeh},
  {McMillan}, {Messina}, {Michalik}, {Millar}, {Mints}, {Molina}, {Molinaro},
  {Moln{\'a}r}, {Montegriffo}, {Mor}, {Morbidelli}, {Morel}, {Morris},
  {Mulone}, {Munoz}, {Muraveva}, {Murphy}, {Musella}, {Noval}, {Ord{\'e}novic},
  {Orr{\`u}}, {Osinde}, {Pagani}, {Pagano}, {Palaversa}, {Palicio}, {Panahi},
  {Pawlak}, {Pe{\~n}alosa Esteller}, {Penttil{\"a}}, {Piersimoni}, {Pineau},
  {Plachy}, {Plum}, {Poggio}, {Poretti}, {Poujoulet}, {Pr{\v{s}}a}, {Pulone},
  {Racero}, {Ragaini}, {Rainer}, {Raiteri}, {Rambaux}, {Ramos}, {Ramos-Lerate},
  {Re Fiorentin}, {Regibo}, {Reyl{\'e}}, {Ripepi}, {Riva}, {Rixon}, {Robichon},
  {Robin}, {Roelens}, {Rohrbasser}, {Romero-G{\'o}mez}, {Rowell}, {Royer},
  {Rybicki}, {Sadowski}, {Sagrist{\`a} Sell{\'e}s}, {Sahlmann}, {Salgado},
  {Salguero}, {Samaras}, {Sanchez Gimenez}, {Sanna}, {Santove{\~n}a},
  {Sarasso}, {Schultheis}, {Sciacca}, {Segol}, {Segovia}, {S{\'e}gransan},
  {Semeux}, {Shahaf}, {Siddiqui}, {Siebert}, {Siltala}, {Slezak}, {Smart},
  {Solano}, {Solitro}, {Souami}, {Souchay}, {Spagna}, {Spoto}, {Steele},
  {Steidelm{\"u}ller}, {Stephenson}, {S{\"u}veges}, {Szabados}, {Szegedi-Elek},
  {Taris}, {Tauran}, {Taylor}, {Teixeira}, {Thuillot}, {Tonello}, {Torra},
  {Torra}, {Turon}, {Unger}, {Vaillant}, {van Dillen}, {Vanel}, {Vecchiato},
  {Viala}, {Vicente}, {Voutsinas}, {Weiler}, {Wevers}, {Wyrzykowski}, {Yoldas},
  {Yvard}, {Zhao}, {Zorec}, {Zucker}, {Zurbach} \& {Zwitter}}]{gaia2021}
{Gaia Collaboration}, {Brown}, A.~G.~A., {Vallenari}, A., {Prusti}, T., {de
  Bruijne}, J.~H.~J., {Babusiaux}, C., {Biermann}, M., {Creevey}, O.~L.,
  {Evans}, D.~W., {Eyer}, L., {Hutton}, A., {Jansen}, F., {Jordi}, C.,
  {Klioner}, S.~A., {Lammers}, U., {Lindegren}, L., {Luri}, X., {Mignard}, F.,
  {Panem}, C., {Pourbaix}, D., {Randich}, S., {Sartoretti}, P., {Soubiran}, C.,
  {Walton}, N.~A., {Arenou}, F., {Bailer-Jones}, C.~A.~L., {Bastian}, U.,
  {Cropper}, M., {Drimmel}, R., {Katz}, D., {Lattanzi}, M.~G., {van Leeuwen},
  F., {Bakker}, J., {Cacciari}, C., {Casta{\~n}eda}, J., {De Angeli}, F.,
  {Ducourant}, C., {Fabricius}, C., {Fouesneau}, M., {Fr{\'e}mat}, Y.,
  {Guerra}, R., {Guerrier}, A., {Guiraud}, J., {Jean-Antoine Piccolo}, A.,
  {Masana}, E., {Messineo}, R., {Mowlavi}, N., {Nicolas}, C., {Nienartowicz},
  K., {Pailler}, F., {Panuzzo}, P., {Riclet}, F., {Roux}, W., {Seabroke},
  G.~M., {Sordo}, R., {Tanga}, P., {Th{\'e}venin}, F., {Gracia-Abril}, G.,
  {Portell}, J., {Teyssier}, D., {Altmann}, M., {Andrae}, R., {Bellas-Velidis},
  I., {Benson}, K., {Berthier}, J., {Blomme}, R., {Brugaletta}, E., {Burgess},
  P.~W., {Busso}, G., {Carry}, B., {Cellino}, A., {Cheek}, N., {Clementini},
  G., {Damerdji}, Y., {Davidson}, M., {Delchambre}, L., {Dell'Oro}, A.,
  {Fern{\'a}ndez-Hern{\'a}ndez}, J., {Galluccio}, L., {Garc{\'\i}a-Lario}, P.,
  {Garcia-Reinaldos}, M., {Gonz{\'a}lez-N{\'u}{\~n}ez}, J., {Gosset}, E.,
  {Haigron}, R., {Halbwachs}, J.~L., {Hambly}, N.~C., {Harrison}, D.~L.,
  {Hatzidimitriou}, D., {Heiter}, U., {Hern{\'a}ndez}, J., {Hestroffer}, D.,
  {Hodgkin}, S.~T., {Holl}, B., {Jan{\ss}en}, K., {Jevardat de Fombelle}, G.,
  {Jordan}, S., {Krone-Martins}, A., {Lanzafame}, A.~C., {L{\"o}ffler}, W.,
  {Lorca}, A., {Manteiga}, M., {Marchal}, O., {Marrese}, P.~M., {Moitinho}, A.,
  {Mora}, A., {Muinonen}, K., {Osborne}, P., {Pancino}, E., {Pauwels}, T.,
  {Petit}, J.~M., {Recio-Blanco}, A., {Richards}, P.~J., {Riello}, M.,
  {Rimoldini}, L., {Robin}, A.~C., {Roegiers}, T., {Rybizki}, J., {Sarro},
  L.~M., {Siopis}, C., {Smith}, M., {Sozzetti}, A., {Ulla}, A., {Utrilla}, E.,
  {van Leeuwen}, M., {van Reeven}, W., {Abbas}, U., {Abreu Aramburu}, A.,
  {Accart}, S., {Aerts}, C., {Aguado}, J.~J., {Ajaj}, M., {Altavilla}, G.,
  {{\'A}lvarez}, M.~A., {{\'A}lvarez Cid-Fuentes}, J., {Alves}, J., {Anderson},
  R.~I., {Anglada Varela}, E., {Antoja}, T., {Audard}, M., {Baines}, D.,
  {Baker}, S.~G., {Balaguer-N{\'u}{\~n}ez}, L., {Balbinot}, E., {Balog}, Z.,
  {Barache}, C., {Barbato}, D., {Barros}, M., {Barstow}, M.~A.,
  {Bartolom{\'e}}, S., {Bassilana}, J.~L., {Bauchet}, N., {Baudesson-Stella},
  A., {Becciani}, U., {Bellazzini}, M., {Bernet}, M., {Bertone}, S., {Bianchi},
  L., {Blanco-Cuaresma}, S., {Boch}, T., {Bombrun}, A., {Bossini}, D.,
  {Bouquillon}, S., {Bragaglia}, A., {Bramante}, L., {Breedt}, E., {Bressan},
  A., {Brouillet}, N., {Bucciarelli}, B., {Burlacu}, A., {Busonero}, D.,
  {Butkevich}, A.~G., {Buzzi}, R., {Caffau}, E., {Cancelliere}, R.,
  {C{\'a}novas}, H., {Cantat-Gaudin}, T., {Carballo}, R., {Carlucci}, T.,
  {Carnerero}, M.~I., {Carrasco}, J.~M., {Casamiquela}, L., {Castellani}, M.,
  {Castro-Ginard}, A., {Castro Sampol}, P., {Chaoul}, L., {Charlot}, P.,
  {Chemin}, L., {Chiavassa}, A., {Cioni}, M. R.~L., {Comoretto}, G., {Cooper},
  W.~J., {Cornez}, T., {Cowell}, S., {Crifo}, F., {Crosta}, M., {Crowley}, C.,
  {Dafonte}, C., {Dapergolas}, A., {David}, M., {David}, P., {de Laverny}, P.,
  {De Luise}, F., {De March}, R., {De Ridder}, J., {de Souza}, R., {de
  Teodoro}, P., {de Torres}, A., {del Peloso}, E.~F., {del Pozo}, E., {Delbo},
  M., {Delgado}, A., {Delgado}, H.~E., {Delisle}, J.~B., {Di Matteo}, P.,
  {Diakite}, S., {Diener}, C., {Distefano}, E., {Dolding}, C., {Eappachen}, D.,
  {Edvardsson}, B., {Enke}, H., {Esquej}, P., {Fabre}, C., {Fabrizio}, M.,
  {Faigler}, S., {Fedorets}, G., {Fernique}, P., {Fienga}, A., {Figueras}, F.,
  {Fouron}, C., {Fragkoudi}, F., {Fraile}, E., {Franke}, F., {Gai}, M.,
  {Garabato}, D., {Garcia-Gutierrez}, A., {Garc{\'\i}a-Torres}, M., {Garofalo},
  A., {Gavras}, P., {Gerlach}, E., {Geyer}, R., {Giacobbe}, P., {Gilmore}, G.,
  {Girona}, S., {Giuffrida}, G., {Gomel}, R., {Gomez}, A.,
  {Gonzalez-Santamaria}, I., {Gonz{\'a}lez-Vidal}, J.~J., {Granvik}, M.,
  {Guti{\'e}rrez-S{\'a}nchez}, R., {Guy}, L.~P., {Hauser}, M., {Haywood}, M.,
  {Helmi}, A., {Hidalgo}, S.~L., {Hilger}, T., {H{\l}adczuk}, N., {Hobbs}, D.,
  {Holland}, G., {Huckle}, H.~E., {Jasniewicz}, G., {Jonker}, P.~G., {Juaristi
  Campillo}, J., {Julbe}, F., {Karbevska}, L., {Kervella}, P., {Khanna}, S.,
  {Kochoska}, A., {Kontizas}, M., {Kordopatis}, G., {Korn}, A.~J.,
  {Kostrzewa-Rutkowska}, Z., {Kruszy{\'n}ska}, K., {Lambert}, S., {Lanza},
  A.~F., {Lasne}, Y., {Le Campion}, J.~F., {Le Fustec}, Y., {Lebreton}, Y.,
  {Lebzelter}, T., {Leccia}, S., {Leclerc}, N., {Lecoeur-Taibi}, I., {Liao},
  S., {Licata}, E., {Lindstr{\o}m}, E.~P., {Lister}, T.~A., {Livanou}, E.,
  {Lobel}, A., {Madrero Pardo}, P., {Managau}, S., {Mann}, R.~G., {Marchant},
  J.~M., {Marconi}, M., {Marcos Santos}, M.~M.~S., {Marinoni}, S., {Marocco},
  F., {Marshall}, D.~J., {Martin Polo}, L., {Mart{\'\i}n-Fleitas}, J.~M.,
  {Masip}, A., {Massari}, D., {Mastrobuono-Battisti}, A., {Mazeh}, T.,
  {McMillan}, P.~J., {Messina}, S., {Michalik}, D., {Millar}, N.~R., {Mints},
  A., {Molina}, D., {Molinaro}, R., {Moln{\'a}r}, L., {Montegriffo}, P., {Mor},
  R., {Morbidelli}, R., {Morel}, T., {Morris}, D., {Mulone}, A.~F., {Munoz},
  D., {Muraveva}, T., {Murphy}, C.~P., {Musella}, I., {Noval}, L.,
  {Ord{\'e}novic}, C., {Orr{\`u}}, G., {Osinde}, J., {Pagani}, C., {Pagano},
  I., {Palaversa}, L., {Palicio}, P.~A., {Panahi}, A., {Pawlak}, M.,
  {Pe{\~n}alosa Esteller}, X., {Penttil{\"a}}, A., {Piersimoni}, A.~M.,
  {Pineau}, F.~X., {Plachy}, E., {Plum}, G., {Poggio}, E., {Poretti}, E.,
  {Poujoulet}, E., {Pr{\v{s}}a}, A., {Pulone}, L., {Racero}, E., {Ragaini}, S.,
  {Rainer}, M., {Raiteri}, C.~M., {Rambaux}, N., {Ramos}, P., {Ramos-Lerate},
  M., {Re Fiorentin}, P., {Regibo}, S., {Reyl{\'e}}, C., {Ripepi}, V., {Riva},
  A., {Rixon}, G., {Robichon}, N., {Robin}, C., {Roelens}, M., {Rohrbasser},
  L., {Romero-G{\'o}mez}, M., {Rowell}, N., {Royer}, F., {Rybicki}, K.~A.,
  {Sadowski}, G., {Sagrist{\`a} Sell{\'e}s}, A., {Sahlmann}, J., {Salgado}, J.,
  {Salguero}, E., {Samaras}, N., {Sanchez Gimenez}, V., {Sanna}, N.,
  {Santove{\~n}a}, R., {Sarasso}, M., {Schultheis}, M., {Sciacca}, E., {Segol},
  M., {Segovia}, J.~C., {S{\'e}gransan}, D., {Semeux}, D., {Shahaf}, S.,
  {Siddiqui}, H.~I., {Siebert}, A., {Siltala}, L., {Slezak}, E., {Smart},
  R.~L., {Solano}, E., {Solitro}, F., {Souami}, D., {Souchay}, J., {Spagna},
  A., {Spoto}, F., {Steele}, I.~A., {Steidelm{\"u}ller}, H., {Stephenson},
  C.~A., {S{\"u}veges}, M., {Szabados}, L., {Szegedi-Elek}, E., {Taris}, F.,
  {Tauran}, G., {Taylor}, M.~B., {Teixeira}, R., {Thuillot}, W., {Tonello}, N.,
  {Torra}, F., {Torra}, J., {Turon}, C., {Unger}, N., {Vaillant}, M., {van
  Dillen}, E., {Vanel}, O., {Vecchiato}, A., {Viala}, Y., {Vicente}, D.,
  {Voutsinas}, S., {Weiler}, M., {Wevers}, T., {Wyrzykowski}, {\L}., {Yoldas},
  A., {Yvard}, P., {Zhao}, H., {Zorec}, J., {Zucker}, S., {Zurbach}, C. \&
  {Zwitter}, T. [2021]  \emph{\aap} \textbf{649}, A1,
  \doi{10.1051/0004-6361/202039657}.

\bibitem[{{Gaia Collaboration} \emph{et~al.}(2016){Gaia Collaboration},
  {Prusti}, {de Bruijne}, {Brown}, {Vallenari}, {Babusiaux}, {Bailer-Jones},
  {Bastian}, {Biermann}, {Evans}, {Eyer}, {Jansen}, {Jordi}, {Klioner},
  {Lammers}, {Lindegren}, {Luri}, {Mignard}, {Milligan}, {Panem}, {Poinsignon},
  {Pourbaix}, {Randich}, {Sarri}, {Sartoretti}, {Siddiqui}, {Soubiran},
  {Valette}, {van Leeuwen}, {Walton}, {Aerts}, {Arenou}, {Cropper}, {Drimmel},
  {H{\o}g}, {Katz}, {Lattanzi}, {O'Mullane}, {Grebel}, {Holland}, {Huc},
  {Passot}, {Bramante}, {Cacciari}, {Casta{\~n}eda}, {Chaoul}, {Cheek}, {De
  Angeli}, {Fabricius}, {Guerra}, {Hern{\'a}ndez}, {Jean-Antoine-Piccolo},
  {Masana}, {Messineo}, {Mowlavi}, {Nienartowicz}, {Ord{\'o}{\~n}ez-Blanco},
  {Panuzzo}, {Portell}, {Richards}, {Riello}, {Seabroke}, {Tanga},
  {Th{\'e}venin}, {Torra}, {Els}, {Gracia-Abril}, {Comoretto},
  {Garcia-Reinaldos}, {Lock}, {Mercier}, {Altmann}, {Andrae}, {Astraatmadja},
  {Bellas-Velidis}, {Benson}, {Berthier}, {Blomme}, {Busso}, {Carry},
  {Cellino}, {Clementini}, {Cowell}, {Creevey}, {Cuypers}, {Davidson}, {De
  Ridder}, {de Torres}, {Delchambre}, {Dell'Oro}, {Ducourant}, {Fr{\'e}mat},
  {Garc{\'\i}a-Torres}, {Gosset}, {Halbwachs}, {Hambly}, {Harrison}, {Hauser},
  {Hestroffer}, {Hodgkin}, {Huckle}, {Hutton}, {Jasniewicz}, {Jordan},
  {Kontizas}, {Korn}, {Lanzafame}, {Manteiga}, {Moitinho}, {Muinonen},
  {Osinde}, {Pancino}, {Pauwels}, {Petit}, {Recio-Blanco}, {Robin}, {Sarro},
  {Siopis}, {Smith}, {Smith}, {Sozzetti}, {Thuillot}, {van Reeven}, {Viala},
  {Abbas}, {Abreu Aramburu}, {Accart}, {Aguado}, {Allan}, {Allasia},
  {Altavilla}, {{\'A}lvarez}, {Alves}, {Anderson}, {Andrei}, {Anglada Varela},
  {Antiche}, {Antoja}, {Ant{\'o}n}, {Arcay}, {Atzei}, {Ayache}, {Bach},
  {Baker}, {Balaguer-N{\'u}{\~n}ez}, {Barache}, {Barata}, {Barbier}, {Barblan},
  {Baroni}, {Barrado y Navascu{\'e}s}, {Barros}, {Barstow}, {Becciani},
  {Bellazzini}, {Bellei}, {Bello Garc{\'\i}a}, {Belokurov}, {Bendjoya},
  {Berihuete}, {Bianchi}, {Bienaym{\'e}}, {Billebaud}, {Blagorodnova},
  {Blanco-Cuaresma}, {Boch}, {Bombrun}, {Borrachero}, {Bouquillon}, {Bourda},
  {Bouy}, {Bragaglia}, {Breddels}, {Brouillet}, {Br{\"u}semeister},
  {Bucciarelli}, {Budnik}, {Burgess}, {Burgon}, {Burlacu}, {Busonero}, {Buzzi},
  {Caffau}, {Cambras}, {Campbell}, {Cancelliere}, {Cantat-Gaudin}, {Carlucci},
  {Carrasco}, {Castellani}, {Charlot}, {Charnas}, {Charvet}, {Chassat},
  {Chiavassa}, {Clotet}, {Cocozza}, {Collins}, {Collins}, {Costigan}, {Crifo},
  {Cross}, {Crosta}, {Crowley}, {Dafonte}, {Damerdji}, {Dapergolas}, {David},
  {David}, {De Cat}, {de Felice}, {de Laverny}, {De Luise}, {De March}, {de
  Martino}, {de Souza}, {Debosscher}, {del Pozo}, {Delbo}, {Delgado},
  {Delgado}, {di Marco}, {Di Matteo}, {Diakite}, {Distefano}, {Dolding}, {Dos
  Anjos}, {Drazinos}, {Dur{\'a}n}, {Dzigan}, {Ecale}, {Edvardsson}, {Enke},
  {Erdmann}, {Escolar}, {Espina}, {Evans}, {Eynard Bontemps}, {Fabre},
  {Fabrizio}, {Faigler}, {Falc{\~a}o}, {Farr{\`a}s Casas}, {Faye}, {Federici},
  {Fedorets}, {Fern{\'a}ndez-Hern{\'a}ndez}, {Fernique}, {Fienga}, {Figueras},
  {Filippi}, {Findeisen}, {Fonti}, {Fouesneau}, {Fraile}, {Fraser}, {Fuchs},
  {Furnell}, {Gai}, {Galleti}, {Galluccio}, {Garabato}, {Garc{\'\i}a-Sedano},
  {Gar{\'e}}, {Garofalo}, {Garralda}, {Gavras}, {Gerssen}, {Geyer}, {Gilmore},
  {Girona}, {Giuffrida}, {Gomes}, {Gonz{\'a}lez-Marcos},
  {Gonz{\'a}lez-N{\'u}{\~n}ez}, {Gonz{\'a}lez-Vidal}, {Granvik}, {Guerrier},
  {Guillout}, {Guiraud}, {G{\'u}rpide}, {Guti{\'e}rrez-S{\'a}nchez}, {Guy},
  {Haigron}, {Hatzidimitriou}, {Haywood}, {Heiter}, {Helmi}, {Hobbs},
  {Hofmann}, {Holl}, {Holland}, {Hunt}, {Hypki}, {Icardi}, {Irwin}, {Jevardat
  de Fombelle}, {Jofr{\'e}}, {Jonker}, {Jorissen}, {Julbe}, {Karampelas},
  {Kochoska}, {Kohley}, {Kolenberg}, {Kontizas}, {Koposov}, {Kordopatis},
  {Koubsky}, {Kowalczyk}, {Krone-Martins}, {Kudryashova}, {Kull}, {Bachchan},
  {Lacoste-Seris}, {Lanza}, {Lavigne}, {Le Poncin-Lafitte}, {Lebreton},
  {Lebzelter}, {Leccia}, {Leclerc}, {Lecoeur-Taibi}, {Lemaitre}, {Lenhardt},
  {Leroux}, {Liao}, {Licata}, {Lindstr{\o}m}, {Lister}, {Livanou}, {Lobel},
  {L{\"o}ffler}, {L{\'o}pez}, {Lopez-Lozano}, {Lorenz}, {Loureiro},
  {MacDonald}, {Magalh{\~a}es Fernandes}, {Managau}, {Mann}, {Mantelet},
  {Marchal}, {Marchant}, {Marconi}, {Marie}, {Marinoni}, {Marrese},
  {Marschalk{\'o}}, {Marshall}, {Mart{\'\i}n-Fleitas}, {Martino}, {Mary},
  {Matijevi{\v{c}}}, {Mazeh}, {McMillan}, {Messina}, {Mestre}, {Michalik},
  {Millar}, {Miranda}, {Molina}, {Molinaro}, {Molinaro}, {Moln{\'a}r},
  {Moniez}, {Montegriffo}, {Monteiro}, {Mor}, {Mora}, {Morbidelli}, {Morel},
  {Morgenthaler}, {Morley}, {Morris}, {Mulone}, {Muraveva}, {Musella},
  {Narbonne}, {Nelemans}, {Nicastro}, {Noval}, {Ord{\'e}novic},
  {Ordieres-Mer{\'e}}, {Osborne}, {Pagani}, {Pagano}, {Pailler}, {Palacin},
  {Palaversa}, {Parsons}, {Paulsen}, {Pecoraro}, {Pedrosa}, {Pentik{\"a}inen},
  {Pereira}, {Pichon}, {Piersimoni}, {Pineau}, {Plachy}, {Plum}, {Poujoulet},
  {Pr{\v{s}}a}, {Pulone}, {Ragaini}, {Rago}, {Rambaux}, {Ramos-Lerate},
  {Ranalli}, {Rauw}, {Read}, {Regibo}, {Renk}, {Reyl{\'e}}, {Ribeiro},
  {Rimoldini}, {Ripepi}, {Riva}, {Rixon}, {Roelens}, {Romero-G{\'o}mez},
  {Rowell}, {Royer}, {Rudolph}, {Ruiz-Dern}, {Sadowski}, {Sagrist{\`a}
  Sell{\'e}s}, {Sahlmann}, {Salgado}, {Salguero}, {Sarasso}, {Savietto},
  {Schnorhk}, {Schultheis}, {Sciacca}, {Segol}, {Segovia}, {Segransan},
  {Serpell}, {Shih}, {Smareglia}, {Smart}, {Smith}, {Solano}, {Solitro},
  {Sordo}, {Soria Nieto}, {Souchay}, {Spagna}, {Spoto}, {Stampa}, {Steele},
  {Steidelm{\"u}ller}, {Stephenson}, {Stoev}, {Suess}, {S{\"u}veges}, {Surdej},
  {Szabados}, {Szegedi-Elek}, {Tapiador}, {Taris}, {Tauran}, {Taylor},
  {Teixeira}, {Terrett}, {Tingley}, {Trager}, {Turon}, {Ulla}, {Utrilla},
  {Valentini}, {van Elteren}, {Van Hemelryck}, {van Leeuwen}, {Varadi},
  {Vecchiato}, {Veljanoski}, {Via}, {Vicente}, {Vogt}, {Voss}, {Votruba},
  {Voutsinas}, {Walmsley}, {Weiler}, {Weingrill}, {Werner}, {Wevers},
  {Whitehead}, {Wyrzykowski}, {Yoldas}, {{\v{Z}}erjal}, {Zucker}, {Zurbach},
  {Zwitter}, {Alecu}, {Allen}, {Allende Prieto}, {Amorim},
  {Anglada-Escud{\'e}}, {Arsenijevic}, {Azaz}, {Balm}, {Beck}, {Bernstein},
  {Bigot}, {Bijaoui}, {Blasco}, {Bonfigli}, {Bono}, {Boudreault}, {Bressan},
  {Brown}, {Brunet}, {Bunclark}, {Buonanno}, {Butkevich}, {Carret}, {Carrion},
  {Chemin}, {Ch{\'e}reau}, {Corcione}, {Darmigny}, {de Boer}, {de Teodoro}, {de
  Zeeuw}, {Delle Luche}, {Domingues}, {Dubath}, {Fodor}, {Fr{\'e}zouls},
  {Fries}, {Fustes}, {Fyfe}, {Gallardo}, {Gallegos}, {Gardiol}, {Gebran},
  {Gomboc}, {G{\'o}mez}, {Grux}, {Gueguen}, {Heyrovsky}, {Hoar}, {Iannicola},
  {Isasi Parache}, {Janotto}, {Joliet}, {Jonckheere}, {Keil}, {Kim},
  {Klagyivik}, {Klar}, {Knude}, {Kochukhov}, {Kolka}, {Kos}, {Kutka}, {Lainey},
  {LeBouquin}, {Liu}, {Loreggia}, {Makarov}, {Marseille}, {Martayan},
  {Martinez-Rubi}, {Massart}, {Meynadier}, {Mignot}, {Munari}, {Nguyen},
  {Nordlander}, {Ocvirk}, {O'Flaherty}, {Olias Sanz}, {Ortiz}, {Osorio},
  {Oszkiewicz}, {Ouzounis}, {Palmer}, {Park}, {Pasquato}, {Peltzer}, {Peralta},
  {P{\'e}turaud}, {Pieniluoma}, {Pigozzi}, {Poels}, {Prat}, {Prod'homme},
  {Raison}, {Rebordao}, {Risquez}, {Rocca-Volmerange}, {Rosen}, {Ruiz-Fuertes},
  {Russo}, {Sembay}, {Serraller Vizcaino}, {Short}, {Siebert}, {Silva},
  {Sinachopoulos}, {Slezak}, {Soffel}, {Sosnowska}, {Strai{\v{z}}ys}, {ter
  Linden}, {Terrell}, {Theil}, {Tiede}, {Troisi}, {Tsalmantza}, {Tur},
  {Vaccari}, {Vachier}, {Valles}, {Van Hamme}, {Veltz}, {Virtanen}, {Wallut},
  {Wichmann}, {Wilkinson}, {Ziaeepour} \& {Zschocke}}]{gaia_mission}
{Gaia Collaboration}, {Prusti}, T., {de Bruijne}, J.~H.~J., {Brown}, A.~G.~A.,
  {Vallenari}, A., {Babusiaux}, C., {Bailer-Jones}, C.~A.~L., {Bastian}, U.,
  {Biermann}, M., {Evans}, D.~W., {Eyer}, L., {Jansen}, F., {Jordi}, C.,
  {Klioner}, S.~A., {Lammers}, U., {Lindegren}, L., {Luri}, X., {Mignard}, F.,
  {Milligan}, D.~J., {Panem}, C., {Poinsignon}, V., {Pourbaix}, D., {Randich},
  S., {Sarri}, G., {Sartoretti}, P., {Siddiqui}, H.~I., {Soubiran}, C.,
  {Valette}, V., {van Leeuwen}, F., {Walton}, N.~A., {Aerts}, C., {Arenou}, F.,
  {Cropper}, M., {Drimmel}, R., {H{\o}g}, E., {Katz}, D., {Lattanzi}, M.~G.,
  {O'Mullane}, W., {Grebel}, E.~K., {Holland}, A.~D., {Huc}, C., {Passot}, X.,
  {Bramante}, L., {Cacciari}, C., {Casta{\~n}eda}, J., {Chaoul}, L., {Cheek},
  N., {De Angeli}, F., {Fabricius}, C., {Guerra}, R., {Hern{\'a}ndez}, J.,
  {Jean-Antoine-Piccolo}, A., {Masana}, E., {Messineo}, R., {Mowlavi}, N.,
  {Nienartowicz}, K., {Ord{\'o}{\~n}ez-Blanco}, D., {Panuzzo}, P., {Portell},
  J., {Richards}, P.~J., {Riello}, M., {Seabroke}, G.~M., {Tanga}, P.,
  {Th{\'e}venin}, F., {Torra}, J., {Els}, S.~G., {Gracia-Abril}, G.,
  {Comoretto}, G., {Garcia-Reinaldos}, M., {Lock}, T., {Mercier}, E.,
  {Altmann}, M., {Andrae}, R., {Astraatmadja}, T.~L., {Bellas-Velidis}, I.,
  {Benson}, K., {Berthier}, J., {Blomme}, R., {Busso}, G., {Carry}, B.,
  {Cellino}, A., {Clementini}, G., {Cowell}, S., {Creevey}, O., {Cuypers}, J.,
  {Davidson}, M., {De Ridder}, J., {de Torres}, A., {Delchambre}, L.,
  {Dell'Oro}, A., {Ducourant}, C., {Fr{\'e}mat}, Y., {Garc{\'\i}a-Torres}, M.,
  {Gosset}, E., {Halbwachs}, J.~L., {Hambly}, N.~C., {Harrison}, D.~L.,
  {Hauser}, M., {Hestroffer}, D., {Hodgkin}, S.~T., {Huckle}, H.~E., {Hutton},
  A., {Jasniewicz}, G., {Jordan}, S., {Kontizas}, M., {Korn}, A.~J.,
  {Lanzafame}, A.~C., {Manteiga}, M., {Moitinho}, A., {Muinonen}, K., {Osinde},
  J., {Pancino}, E., {Pauwels}, T., {Petit}, J.~M., {Recio-Blanco}, A.,
  {Robin}, A.~C., {Sarro}, L.~M., {Siopis}, C., {Smith}, M., {Smith}, K.~W.,
  {Sozzetti}, A., {Thuillot}, W., {van Reeven}, W., {Viala}, Y., {Abbas}, U.,
  {Abreu Aramburu}, A., {Accart}, S., {Aguado}, J.~J., {Allan}, P.~M.,
  {Allasia}, W., {Altavilla}, G., {{\'A}lvarez}, M.~A., {Alves}, J.,
  {Anderson}, R.~I., {Andrei}, A.~H., {Anglada Varela}, E., {Antiche}, E.,
  {Antoja}, T., {Ant{\'o}n}, S., {Arcay}, B., {Atzei}, A., {Ayache}, L.,
  {Bach}, N., {Baker}, S.~G., {Balaguer-N{\'u}{\~n}ez}, L., {Barache}, C.,
  {Barata}, C., {Barbier}, A., {Barblan}, F., {Baroni}, M., {Barrado y
  Navascu{\'e}s}, D., {Barros}, M., {Barstow}, M.~A., {Becciani}, U.,
  {Bellazzini}, M., {Bellei}, G., {Bello Garc{\'\i}a}, A., {Belokurov}, V.,
  {Bendjoya}, P., {Berihuete}, A., {Bianchi}, L., {Bienaym{\'e}}, O.,
  {Billebaud}, F., {Blagorodnova}, N., {Blanco-Cuaresma}, S., {Boch}, T.,
  {Bombrun}, A., {Borrachero}, R., {Bouquillon}, S., {Bourda}, G., {Bouy}, H.,
  {Bragaglia}, A., {Breddels}, M.~A., {Brouillet}, N., {Br{\"u}semeister}, T.,
  {Bucciarelli}, B., {Budnik}, F., {Burgess}, P., {Burgon}, R., {Burlacu}, A.,
  {Busonero}, D., {Buzzi}, R., {Caffau}, E., {Cambras}, J., {Campbell}, H.,
  {Cancelliere}, R., {Cantat-Gaudin}, T., {Carlucci}, T., {Carrasco}, J.~M.,
  {Castellani}, M., {Charlot}, P., {Charnas}, J., {Charvet}, P., {Chassat}, F.,
  {Chiavassa}, A., {Clotet}, M., {Cocozza}, G., {Collins}, R.~S., {Collins},
  P., {Costigan}, G., {Crifo}, F., {Cross}, N.~J.~G., {Crosta}, M., {Crowley},
  C., {Dafonte}, C., {Damerdji}, Y., {Dapergolas}, A., {David}, P., {David},
  M., {De Cat}, P., {de Felice}, F., {de Laverny}, P., {De Luise}, F., {De
  March}, R., {de Martino}, D., {de Souza}, R., {Debosscher}, J., {del Pozo},
  E., {Delbo}, M., {Delgado}, A., {Delgado}, H.~E., {di Marco}, F., {Di
  Matteo}, P., {Diakite}, S., {Distefano}, E., {Dolding}, C., {Dos Anjos}, S.,
  {Drazinos}, P., {Dur{\'a}n}, J., {Dzigan}, Y., {Ecale}, E., {Edvardsson}, B.,
  {Enke}, H., {Erdmann}, M., {Escolar}, D., {Espina}, M., {Evans}, N.~W.,
  {Eynard Bontemps}, G., {Fabre}, C., {Fabrizio}, M., {Faigler}, S.,
  {Falc{\~a}o}, A.~J., {Farr{\`a}s Casas}, M., {Faye}, F., {Federici}, L.,
  {Fedorets}, G., {Fern{\'a}ndez-Hern{\'a}ndez}, J., {Fernique}, P., {Fienga},
  A., {Figueras}, F., {Filippi}, F., {Findeisen}, K., {Fonti}, A., {Fouesneau},
  M., {Fraile}, E., {Fraser}, M., {Fuchs}, J., {Furnell}, R., {Gai}, M.,
  {Galleti}, S., {Galluccio}, L., {Garabato}, D., {Garc{\'\i}a-Sedano}, F.,
  {Gar{\'e}}, P., {Garofalo}, A., {Garralda}, N., {Gavras}, P., {Gerssen}, J.,
  {Geyer}, R., {Gilmore}, G., {Girona}, S., {Giuffrida}, G., {Gomes}, M.,
  {Gonz{\'a}lez-Marcos}, A., {Gonz{\'a}lez-N{\'u}{\~n}ez}, J.,
  {Gonz{\'a}lez-Vidal}, J.~J., {Granvik}, M., {Guerrier}, A., {Guillout}, P.,
  {Guiraud}, J., {G{\'u}rpide}, A., {Guti{\'e}rrez-S{\'a}nchez}, R., {Guy},
  L.~P., {Haigron}, R., {Hatzidimitriou}, D., {Haywood}, M., {Heiter}, U.,
  {Helmi}, A., {Hobbs}, D., {Hofmann}, W., {Holl}, B., {Holland}, G., {Hunt},
  J.~A.~S., {Hypki}, A., {Icardi}, V., {Irwin}, M., {Jevardat de Fombelle}, G.,
  {Jofr{\'e}}, P., {Jonker}, P.~G., {Jorissen}, A., {Julbe}, F., {Karampelas},
  A., {Kochoska}, A., {Kohley}, R., {Kolenberg}, K., {Kontizas}, E., {Koposov},
  S.~E., {Kordopatis}, G., {Koubsky}, P., {Kowalczyk}, A., {Krone-Martins}, A.,
  {Kudryashova}, M., {Kull}, I., {Bachchan}, R.~K., {Lacoste-Seris}, F.,
  {Lanza}, A.~F., {Lavigne}, J.~B., {Le Poncin-Lafitte}, C., {Lebreton}, Y.,
  {Lebzelter}, T., {Leccia}, S., {Leclerc}, N., {Lecoeur-Taibi}, I.,
  {Lemaitre}, V., {Lenhardt}, H., {Leroux}, F., {Liao}, S., {Licata}, E.,
  {Lindstr{\o}m}, H.~E.~P., {Lister}, T.~A., {Livanou}, E., {Lobel}, A.,
  {L{\"o}ffler}, W., {L{\'o}pez}, M., {Lopez-Lozano}, A., {Lorenz}, D.,
  {Loureiro}, T., {MacDonald}, I., {Magalh{\~a}es Fernandes}, T., {Managau},
  S., {Mann}, R.~G., {Mantelet}, G., {Marchal}, O., {Marchant}, J.~M.,
  {Marconi}, M., {Marie}, J., {Marinoni}, S., {Marrese}, P.~M.,
  {Marschalk{\'o}}, G., {Marshall}, D.~J., {Mart{\'\i}n-Fleitas}, J.~M.,
  {Martino}, M., {Mary}, N., {Matijevi{\v{c}}}, G., {Mazeh}, T., {McMillan},
  P.~J., {Messina}, S., {Mestre}, A., {Michalik}, D., {Millar}, N.~R.,
  {Miranda}, B.~M.~H., {Molina}, D., {Molinaro}, R., {Molinaro}, M.,
  {Moln{\'a}r}, L., {Moniez}, M., {Montegriffo}, P., {Monteiro}, D., {Mor}, R.,
  {Mora}, A., {Morbidelli}, R., {Morel}, T., {Morgenthaler}, S., {Morley}, T.,
  {Morris}, D., {Mulone}, A.~F., {Muraveva}, T., {Musella}, I., {Narbonne}, J.,
  {Nelemans}, G., {Nicastro}, L., {Noval}, L., {Ord{\'e}novic}, C.,
  {Ordieres-Mer{\'e}}, J., {Osborne}, P., {Pagani}, C., {Pagano}, I.,
  {Pailler}, F., {Palacin}, H., {Palaversa}, L., {Parsons}, P., {Paulsen}, T.,
  {Pecoraro}, M., {Pedrosa}, R., {Pentik{\"a}inen}, H., {Pereira}, J.,
  {Pichon}, B., {Piersimoni}, A.~M., {Pineau}, F.~X., {Plachy}, E., {Plum}, G.,
  {Poujoulet}, E., {Pr{\v{s}}a}, A., {Pulone}, L., {Ragaini}, S., {Rago}, S.,
  {Rambaux}, N., {Ramos-Lerate}, M., {Ranalli}, P., {Rauw}, G., {Read}, A.,
  {Regibo}, S., {Renk}, F., {Reyl{\'e}}, C., {Ribeiro}, R.~A., {Rimoldini}, L.,
  {Ripepi}, V., {Riva}, A., {Rixon}, G., {Roelens}, M., {Romero-G{\'o}mez}, M.,
  {Rowell}, N., {Royer}, F., {Rudolph}, A., {Ruiz-Dern}, L., {Sadowski}, G.,
  {Sagrist{\`a} Sell{\'e}s}, T., {Sahlmann}, J., {Salgado}, J., {Salguero}, E.,
  {Sarasso}, M., {Savietto}, H., {Schnorhk}, A., {Schultheis}, M., {Sciacca},
  E., {Segol}, M., {Segovia}, J.~C., {Segransan}, D., {Serpell}, E., {Shih},
  I.~C., {Smareglia}, R., {Smart}, R.~L., {Smith}, C., {Solano}, E., {Solitro},
  F., {Sordo}, R., {Soria Nieto}, S., {Souchay}, J., {Spagna}, A., {Spoto}, F.,
  {Stampa}, U., {Steele}, I.~A., {Steidelm{\"u}ller}, H., {Stephenson}, C.~A.,
  {Stoev}, H., {Suess}, F.~F., {S{\"u}veges}, M., {Surdej}, J., {Szabados}, L.,
  {Szegedi-Elek}, E., {Tapiador}, D., {Taris}, F., {Tauran}, G., {Taylor},
  M.~B., {Teixeira}, R., {Terrett}, D., {Tingley}, B., {Trager}, S.~C.,
  {Turon}, C., {Ulla}, A., {Utrilla}, E., {Valentini}, G., {van Elteren}, A.,
  {Van Hemelryck}, E., {van Leeuwen}, M., {Varadi}, M., {Vecchiato}, A.,
  {Veljanoski}, J., {Via}, T., {Vicente}, D., {Vogt}, S., {Voss}, H.,
  {Votruba}, V., {Voutsinas}, S., {Walmsley}, G., {Weiler}, M., {Weingrill},
  K., {Werner}, D., {Wevers}, T., {Whitehead}, G., {Wyrzykowski}, {\L}.,
  {Yoldas}, A., {{\v{Z}}erjal}, M., {Zucker}, S., {Zurbach}, C., {Zwitter}, T.,
  {Alecu}, A., {Allen}, M., {Allende Prieto}, C., {Amorim}, A.,
  {Anglada-Escud{\'e}}, G., {Arsenijevic}, V., {Azaz}, S., {Balm}, P., {Beck},
  M., {Bernstein}, H.~H., {Bigot}, L., {Bijaoui}, A., {Blasco}, C., {Bonfigli},
  M., {Bono}, G., {Boudreault}, S., {Bressan}, A., {Brown}, S., {Brunet},
  P.~M., {Bunclark}, P., {Buonanno}, R., {Butkevich}, A.~G., {Carret}, C.,
  {Carrion}, C., {Chemin}, L., {Ch{\'e}reau}, F., {Corcione}, L., {Darmigny},
  E., {de Boer}, K.~S., {de Teodoro}, P., {de Zeeuw}, P.~T., {Delle Luche}, C.,
  {Domingues}, C.~D., {Dubath}, P., {Fodor}, F., {Fr{\'e}zouls}, B., {Fries},
  A., {Fustes}, D., {Fyfe}, D., {Gallardo}, E., {Gallegos}, J., {Gardiol}, D.,
  {Gebran}, M., {Gomboc}, A., {G{\'o}mez}, A., {Grux}, E., {Gueguen}, A.,
  {Heyrovsky}, A., {Hoar}, J., {Iannicola}, G., {Isasi Parache}, Y., {Janotto},
  A.~M., {Joliet}, E., {Jonckheere}, A., {Keil}, R., {Kim}, D.~W., {Klagyivik},
  P., {Klar}, J., {Knude}, J., {Kochukhov}, O., {Kolka}, I., {Kos}, J.,
  {Kutka}, A., {Lainey}, V., {LeBouquin}, D., {Liu}, C., {Loreggia}, D.,
  {Makarov}, V.~V., {Marseille}, M.~G., {Martayan}, C., {Martinez-Rubi}, O.,
  {Massart}, B., {Meynadier}, F., {Mignot}, S., {Munari}, U., {Nguyen}, A.~T.,
  {Nordlander}, T., {Ocvirk}, P., {O'Flaherty}, K.~S., {Olias Sanz}, A.,
  {Ortiz}, P., {Osorio}, J., {Oszkiewicz}, D., {Ouzounis}, A., {Palmer}, M.,
  {Park}, P., {Pasquato}, E., {Peltzer}, C., {Peralta}, J., {P{\'e}turaud}, F.,
  {Pieniluoma}, T., {Pigozzi}, E., {Poels}, J., {Prat}, G., {Prod'homme}, T.,
  {Raison}, F., {Rebordao}, J.~M., {Risquez}, D., {Rocca-Volmerange}, B.,
  {Rosen}, S., {Ruiz-Fuertes}, M.~I., {Russo}, F., {Sembay}, S., {Serraller
  Vizcaino}, I., {Short}, A., {Siebert}, A., {Silva}, H., {Sinachopoulos}, D.,
  {Slezak}, E., {Soffel}, M., {Sosnowska}, D., {Strai{\v{z}}ys}, V., {ter
  Linden}, M., {Terrell}, D., {Theil}, S., {Tiede}, C., {Troisi}, L.,
  {Tsalmantza}, P., {Tur}, D., {Vaccari}, M., {Vachier}, F., {Valles}, P., {Van
  Hamme}, W., {Veltz}, L., {Virtanen}, J., {Wallut}, J.~M., {Wichmann}, R.,
  {Wilkinson}, M.~I., {Ziaeepour}, H. \& {Zschocke}, S. [2016]  \emph{\aap}
  \textbf{595}, A1, \doi{10.1051/0004-6361/201629272}.

\bibitem[{{Gunn} \emph{et~al.}(2006){Gunn}, {Siegmund}, {Mannery}, {Owen},
  {Hull}, {Leger}, {Carey}, {Knapp}, {York}, {Boroski}, {Kent}, {Lupton},
  {Rockosi}, {Evans}, {Waddell}, {Anderson}, {Annis}, {Barentine}, {Bartoszek},
  {Bastian}, {Bracker}, {Brewington}, {Briegel}, {Brinkmann}, {Brown}, {Carr},
  {Czarapata}, {Drennan}, {Dombeck}, {Federwitz}, {Gillespie}, {Gonzales},
  {Hansen}, {Harvanek}, {Hayes}, {Jordan}, {Kinney}, {Klaene}, {Kleinman},
  {Kron}, {Kresinski}, {Lee}, {Limmongkol}, {Lindenmeyer}, {Long}, {Loomis},
  {McGehee}, {Mantsch}, {Neilsen}, {Neswold}, {Newman}, {Nitta}, {Peoples},
  {Pier}, {Prieto}, {Prosapio}, {Rivetta}, {Schneider}, {Snedden} \&
  {Wang}}]{sdss_telescope}
{Gunn}, J.~E., {Siegmund}, W.~A., {Mannery}, E.~J., {Owen}, R.~E., {Hull},
  C.~L., {Leger}, R.~F., {Carey}, L.~N., {Knapp}, G.~R., {York}, D.~G.,
  {Boroski}, W.~N., {Kent}, S.~M., {Lupton}, R.~H., {Rockosi}, C.~M., {Evans},
  M.~L., {Waddell}, P., {Anderson}, J.~E., {Annis}, J., {Barentine}, J.~C.,
  {Bartoszek}, L.~M., {Bastian}, S., {Bracker}, S.~B., {Brewington}, H.~J.,
  {Briegel}, C.~I., {Brinkmann}, J., {Brown}, Y.~J., {Carr}, M.~A.,
  {Czarapata}, P.~C., {Drennan}, C.~C., {Dombeck}, T., {Federwitz}, G.~R.,
  {Gillespie}, B.~A., {Gonzales}, C., {Hansen}, S.~U., {Harvanek}, M., {Hayes},
  J., {Jordan}, W., {Kinney}, E., {Klaene}, M., {Kleinman}, S.~J., {Kron},
  R.~G., {Kresinski}, J., {Lee}, G., {Limmongkol}, S., {Lindenmeyer}, C.~W.,
  {Long}, D.~C., {Loomis}, C.~L., {McGehee}, P.~M., {Mantsch}, P.~M.,
  {Neilsen}, J., Eric~H., {Neswold}, R.~M., {Newman}, P.~R., {Nitta}, A.,
  {Peoples}, J., John, {Pier}, J.~R., {Prieto}, P.~S., {Prosapio}, A.,
  {Rivetta}, C., {Schneider}, D.~P., {Snedden}, S. \& {Wang}, S.-i. [2006]
  \emph{\aj} \textbf{131},  2332, \doi{10.1086/500975}.

\bibitem[{{Jordi} \emph{et~al.}(2010){Jordi}, {Gebran}, {Carrasco}, {de
  Bruijne}, {Voss}, {Fabricius}, {Knude}, {Vallenari}, {Kohley} \&
  {Mora}}]{jordi}
{Jordi}, C., {Gebran}, M., {Carrasco}, J.~M., {de Bruijne}, J., {Voss}, H.,
  {Fabricius}, C., {Knude}, J., {Vallenari}, A., {Kohley}, R. \& {Mora}, A.
  [2010]  \emph{\aap} \textbf{523}, A48, \doi{10.1051/0004-6361/201015441}.

\bibitem[{{Kumar} \emph{et~al.}(2022){Kumar}, {Negi}, {Ailawadhi}, {Mishra},
  {Pradhan}, {Misra}, {Hickson} \& {Surdej}}]{ILMT_pipeline}
{Kumar}, B., {Negi}, V., {Ailawadhi}, B., {Mishra}, S., {Pradhan}, B., {Misra},
  K., {Hickson}, P. \& {Surdej}, J. [2022]  \emph{Journal of Astrophysics and
  Astronomy} \textbf{43}, 10, \doi{10.1007/s12036-021-09795-3}.

\bibitem[{{Landolt}(1992)}]{landolt1992}
{Landolt}, A.~U. [1992]  \emph{\aj} \textbf{104},  340, \doi{10.1086/116242}.

\bibitem[{{Lang} \emph{et~al.}(2010){Lang}, {Hogg}, {Mierle}, {Blanton} \&
  {Roweis}}]{astrometry.net}
{Lang}, D., {Hogg}, D.~W., {Mierle}, K., {Blanton}, M. \& {Roweis}, S. [2010]
  \emph{\aj} \textbf{139},  1782, \doi{10.1088/0004-6256/139/5/1782}.

\bibitem[{{Lindegren} \emph{et~al.}(2021){Lindegren}, {Klioner},
  {Hern{\'a}ndez}, {Bombrun}, {Ramos-Lerate}, {Steidelm{\"u}ller}, {Bastian},
  {Biermann}, {de Torres}, {Gerlach}, {Geyer}, {Hilger}, {Hobbs}, {Lammers},
  {McMillan}, {Stephenson}, {Casta{\~n}eda}, {Davidson}, {Fabricius},
  {Gracia-Abril}, {Portell}, {Rowell}, {Teyssier}, {Torra}, {Bartolom{\'e}},
  {Clotet}, {Garralda}, {Gonz{\'a}lez-Vidal}, {Torra}, {Abbas}, {Altmann},
  {Anglada Varela}, {Balaguer-N{\'u}{\~n}ez}, {Balog}, {Barache}, {Becciani},
  {Bernet}, {Bertone}, {Bianchi}, {Bouquillon}, {Brown}, {Bucciarelli},
  {Busonero}, {Butkevich}, {Buzzi}, {Cancelliere}, {Carlucci}, {Charlot},
  {Cioni}, {Crosta}, {Crowley}, {del Peloso}, {del Pozo}, {Drimmel}, {Esquej},
  {Fienga}, {Fraile}, {Gai}, {Garcia-Reinaldos}, {Guerra}, {Hambly}, {Hauser},
  {Jan{\ss}en}, {Jordan}, {Kostrzewa-Rutkowska}, {Lattanzi}, {Liao}, {Licata},
  {Lister}, {L{\"o}ffler}, {Marchant}, {Masip}, {Mignard}, {Mints}, {Molina},
  {Mora}, {Morbidelli}, {Murphy}, {Pagani}, {Panuzzo}, {Pe{\~n}alosa Esteller},
  {Poggio}, {Re Fiorentin}, {Riva}, {Sagrist{\`a} Sell{\'e}s}, {Sanchez
  Gimenez}, {Sarasso}, {Sciacca}, {Siddiqui}, {Smart}, {Souami}, {Spagna},
  {Steele}, {Taris}, {Utrilla}, {van Reeven} \& {Vecchiato}}]{lindegren2021}
{Lindegren}, L., {Klioner}, S.~A., {Hern{\'a}ndez}, J., {Bombrun}, A.,
  {Ramos-Lerate}, M., {Steidelm{\"u}ller}, H., {Bastian}, U., {Biermann}, M.,
  {de Torres}, A., {Gerlach}, E., {Geyer}, R., {Hilger}, T., {Hobbs}, D.,
  {Lammers}, U., {McMillan}, P.~J., {Stephenson}, C.~A., {Casta{\~n}eda}, J.,
  {Davidson}, M., {Fabricius}, C., {Gracia-Abril}, G., {Portell}, J., {Rowell},
  N., {Teyssier}, D., {Torra}, F., {Bartolom{\'e}}, S., {Clotet}, M.,
  {Garralda}, N., {Gonz{\'a}lez-Vidal}, J.~J., {Torra}, J., {Abbas}, U.,
  {Altmann}, M., {Anglada Varela}, E., {Balaguer-N{\'u}{\~n}ez}, L., {Balog},
  Z., {Barache}, C., {Becciani}, U., {Bernet}, M., {Bertone}, S., {Bianchi},
  L., {Bouquillon}, S., {Brown}, A.~G.~A., {Bucciarelli}, B., {Busonero}, D.,
  {Butkevich}, A.~G., {Buzzi}, R., {Cancelliere}, R., {Carlucci}, T.,
  {Charlot}, P., {Cioni}, M. R.~L., {Crosta}, M., {Crowley}, C., {del Peloso},
  E.~F., {del Pozo}, E., {Drimmel}, R., {Esquej}, P., {Fienga}, A., {Fraile},
  E., {Gai}, M., {Garcia-Reinaldos}, M., {Guerra}, R., {Hambly}, N.~C.,
  {Hauser}, M., {Jan{\ss}en}, K., {Jordan}, S., {Kostrzewa-Rutkowska}, Z.,
  {Lattanzi}, M.~G., {Liao}, S., {Licata}, E., {Lister}, T.~A., {L{\"o}ffler},
  W., {Marchant}, J.~M., {Masip}, A., {Mignard}, F., {Mints}, A., {Molina}, D.,
  {Mora}, A., {Morbidelli}, R., {Murphy}, C.~P., {Pagani}, C., {Panuzzo}, P.,
  {Pe{\~n}alosa Esteller}, X., {Poggio}, E., {Re Fiorentin}, P., {Riva}, A.,
  {Sagrist{\`a} Sell{\'e}s}, A., {Sanchez Gimenez}, V., {Sarasso}, M.,
  {Sciacca}, E., {Siddiqui}, H.~I., {Smart}, R.~L., {Souami}, D., {Spagna}, A.,
  {Steele}, I.~A., {Taris}, F., {Utrilla}, E., {van Reeven}, W. \& {Vecchiato},
  A. [2021]  \emph{\aap} \textbf{649}, A2, \doi{10.1051/0004-6361/202039709}.

\bibitem[{{Mandal} \emph{et~al.}(2020){Mandal}, {Pradhan}, {Surdej}, {Stalin},
  {Sagar} \& {Mathew}}]{Mandal_2020}
{Mandal}, A.~K., {Pradhan}, B., {Surdej}, J., {Stalin}, C.~S., {Sagar}, R. \&
  {Mathew}, B. [2020]  \emph{Journal of Astrophysics and Astronomy}
  \textbf{41}, 22, \doi{10.1007/s12036-020-09642-x}.

\bibitem[{Marrese \emph{et~al.}(2019)Marrese, Marinoni, Fabrizio \&
  Altavilla}]{Marrese_2019}
Marrese, P.~M., Marinoni, S., Fabrizio, M. \& Altavilla, G. [2019]
  \emph{Astronomy \& Astrophysics} \textbf{621},  A144,
  \doi{10.1051/0004-6361/201834142},
  \urlprefix\url{http://dx.doi.org/10.1051/0004-6361/201834142}.

\bibitem[{{Pineau} \emph{et~al.}(2011){Pineau}, {Motch}, {Carrera}, {Della
  Ceca}, {Derri{\`e}re}, {Michel}, {Schwope} \& {Watson}}]{pineau}
{Pineau}, F.~X., {Motch}, C., {Carrera}, F., {Della Ceca}, R., {Derri{\`e}re},
  S., {Michel}, L., {Schwope}, A. \& {Watson}, M.~G. [2011]  \emph{\aap}
  \textbf{527}, A126, \doi{10.1051/0004-6361/201015141}.

\bibitem[{{Riello} \emph{et~al.}(2021){Riello}, {De Angeli}, {Evans},
  {Montegriffo}, {Carrasco}, {Busso}, {Palaversa}, {Burgess}, {Diener},
  {Davidson}, {Rowell}, {Fabricius}, {Jordi}, {Bellazzini}, {Pancino},
  {Harrison}, {Cacciari}, {van Leeuwen}, {Hambly}, {Hodgkin}, {Osborne},
  {Altavilla}, {Barstow}, {Brown}, {Castellani}, {Cowell}, {De Luise},
  {Gilmore}, {Giuffrida}, {Hidalgo}, {Holland}, {Marinoni}, {Pagani},
  {Piersimoni}, {Pulone}, {Ragaini}, {Rainer}, {Richards}, {Sanna}, {Walton},
  {Weiler} \& {Yoldas}}]{riello2021}
{Riello}, M., {De Angeli}, F., {Evans}, D.~W., {Montegriffo}, P., {Carrasco},
  J.~M., {Busso}, G., {Palaversa}, L., {Burgess}, P.~W., {Diener}, C.,
  {Davidson}, M., {Rowell}, N., {Fabricius}, C., {Jordi}, C., {Bellazzini}, M.,
  {Pancino}, E., {Harrison}, D.~L., {Cacciari}, C., {van Leeuwen}, F.,
  {Hambly}, N.~C., {Hodgkin}, S.~T., {Osborne}, P.~J., {Altavilla}, G.,
  {Barstow}, M.~A., {Brown}, A.~G.~A., {Castellani}, M., {Cowell}, S., {De
  Luise}, F., {Gilmore}, G., {Giuffrida}, G., {Hidalgo}, S., {Holland}, G.,
  {Marinoni}, S., {Pagani}, C., {Piersimoni}, A.~M., {Pulone}, L., {Ragaini},
  S., {Rainer}, M., {Richards}, P.~J., {Sanna}, N., {Walton}, N.~A., {Weiler},
  M. \& {Yoldas}, A. [2021]  \emph{\aap} \textbf{649}, A3,
  \doi{10.1051/0004-6361/202039587}.

\bibitem[{{Sagar} \emph{et~al.}(2012){Sagar}, {Kumar}, {Omar} \&
  {Pandey}}]{Sagar2012}
{Sagar}, R., {Kumar}, B., {Omar}, A. \& {Pandey}, A.~K. [2012]  \enquote{{New
  optical telescope projects at Devasthal Observatory},}  \emph{Ground-based
  and Airborne Telescopes IV}, eds. {Stepp}, L.~M., {Gilmozzi}, R. \& {Hall},
  H.~J., p. 84441T, \doi{10.1117/12.925634}.

\bibitem[{{Sagar} \emph{et~al.}(2000){Sagar}, {Stalin}, {Pandey}, {Uddin},
  {Mohan}, {Sanwal}, {Gupta}, {Yadav}, {Durgapal}, {Joshi}, {Kumar}, {Gupta},
  {Joshi}, {Srivastava}, {Chaubey}, {Singh}, {Pant} \& {Gupta}}]{sagar2000}
{Sagar}, R., {Stalin}, C.~S., {Pandey}, A.~K., {Uddin}, W., {Mohan}, V.,
  {Sanwal}, B.~B., {Gupta}, S.~K., {Yadav}, R.~K.~S., {Durgapal}, A.~K.,
  {Joshi}, S., {Kumar}, B., {Gupta}, A.~C., {Joshi}, Y.~C., {Srivastava},
  J.~B., {Chaubey}, U.~S., {Singh}, M., {Pant}, P. \& {Gupta}, K.~G. [2000]
  \emph{\aaps} \textbf{144},  349, \doi{10.1051/aas:2000213}.

\bibitem[{{Stalin} \emph{et~al.}(2001){Stalin}, {Sagar}, {Pant}, {Mohan},
  {Kumar}, {Joshi}, {Yadav}, {Joshi}, {Chandra}, {Durgapal} \&
  {Uddin}}]{stalin2001}
{Stalin}, C.~S., {Sagar}, R., {Pant}, P., {Mohan}, V., {Kumar}, B., {Joshi},
  Y.~C., {Yadav}, R.~K.~S., {Joshi}, S., {Chandra}, R., {Durgapal}, A.~K. \&
  {Uddin}, W. [2001]  \emph{Bulletin of the Astronomical Society of India}
  \textbf{29},  39.

\bibitem[{{Surdej} \emph{et~al.}(2018){Surdej}, {Hickson}, {Borra}, {Swings},
  {Habraken}, {Akhunov}, {Bartczak}, {Chand}, {De Becker}, {Delchambre},
  {Finet}, {Kumar}, {Pandey}, {Pospieszalska}, {Pradhan}, {Sagar}, {Wertz}, {De
  Cat}, {Denis}, {de Ville}, {Jaiswar}, {Lampens}, {Nanjappa} \&
  {Tortolani}}]{surdej2018}
{Surdej}, J., {Hickson}, P., {Borra}, H., {Swings}, J.-P., {Habraken}, S.,
  {Akhunov}, T., {Bartczak}, P., {Chand}, H., {De Becker}, M., {Delchambre},
  L., {Finet}, F., {Kumar}, B., {Pandey}, A., {Pospieszalska}, A., {Pradhan},
  B., {Sagar}, R., {Wertz}, O., {De Cat}, P., {Denis}, S., {de Ville}, J.,
  {Jaiswar}, M.~K., {Lampens}, P., {Nanjappa}, N. \& {Tortolani}, J.-M. [2018]
  \emph{Bulletin de la Societe Royale des Sciences de Liege} \textbf{87},  68.

\end{thebibliography}
}

\end{document}